\documentclass[iop]{emulateapj}

\accepted{29 April, 2014}

\usepackage{epsfig}
\usepackage{natbib}
\usepackage{mathtools}
\usepackage{graphicx}

\def\simgt{\lower.5ex\hbox{$\; \buildrel > \over \sim \;$}}
\def\simlt{\lower.5ex\hbox{$\; \buildrel < \over \sim \;$}}

\def\amin{\ifmmode^{\prime}\else$^{\prime}$\fi}
\def\asec{\ifmmode^{\prime\prime}\else$^{\prime\prime}$\fi}

\def\simgt{\lower.5ex\hbox{$\; \buildrel > \over \sim \;$}}
\def\simlt{\lower.5ex\hbox{$\; \buildrel < \over \sim \;$}}

\newcommand\chandra{{\it Chandra}}

\newcommand\xmm{{\it XMM-Newton}}
\newcommand\XMM{{\it XMM-Newton}}
\newcommand\integral{{\it INTEGRAL}/IBIS}

\newcommand\nustar{{\it NuSTAR\/}}

\def\g21{G21.5-0.9}

\newcommand\psr{PSR~J1833$-$1034}

\slugcomment{The Astrophysical Journal}

\shorttitle{\nustar\ observations of \g21}
\shortauthors{}

\begin{document}

\title{\nustar\ study of Hard X-Ray Morphology and Spectroscopy of PWN G21.5-0.9 }

\author{Melania Nynka\altaffilmark{1}, Charles J. Hailey\altaffilmark{1}, Stephen P. Reynolds\altaffilmark{2},
Hongjun An\altaffilmark{3}, 
Frederick K. Baganoff\altaffilmark{4},
Steven E. Boggs\altaffilmark{5}, Finn E. Christensen\altaffilmark{6}, William W. Craig\altaffilmark{7,8}, 
Eric V. Gotthelf\altaffilmark{1}, Brian W. Grefenstette\altaffilmark{9}, Fiona A. Harrison\altaffilmark{9}, 
Roman Krivonos\altaffilmark{5}, Kristin K. Madsen\altaffilmark{9}, Kaya Mori\altaffilmark{1}, Kerstin Perez\altaffilmark{1}, 
Daniel Stern\altaffilmark{10}, Daniel R. Wik\altaffilmark{11}, William W. Zhang\altaffilmark{11}, Andreas Zoglauer\altaffilmark{8}
}

\altaffiltext{1}{Columbia Astrophysics Laboratory, Columbia University, New York, NY 10027, USA}
\altaffiltext{2}{Physics Department, NC State University, Raleigh, NC 27695, USA}
\altaffiltext{3}{Department of Physics, McGill University, Rutherford Physics Building, 3600 University Street, Montreal, Quebec H3A 2T8, Canada}
\altaffiltext{4}{Center for Space Research, Massachusetts Institute of Technology, Cambridge, MA 02139-4307}
\altaffiltext{5}{Space Sciences Laboratory, University of California, Berkeley, CA 94720, USA}
\altaffiltext{6}{DTU Space, National Space Institute, Technical University of Denmark, Elektrovej 327, DK-2800 Lyngby, Denmark}
\altaffiltext{7}{Lawrence Livermore National Laboratory, Livermore, CA 94550, USA }
\altaffiltext{8}{Space Sciences Laboratory, University of California, Berkeley, CA 94720, USA}
\altaffiltext{9}{Cahill Center for Astronomy and Astrophysics, California Institute of Technology, Pasadena, CA 91125}
\altaffiltext{10}{Jet Propulsion Laboratory, California Institute of Technology, Pasadena, CA 91109, USA}
\altaffiltext{11}{NASA Goddard Space Flight Center, Greenbelt, MD 20771, USA}

\begin{abstract}

We present \nustar\ high energy X-ray observations of the pulsar wind nebula (PWN)/supernova remnant \g21.   We detect integrated emission from the nebula up to 
$\sim40$~keV, and  resolve individual spatial features over a broad X-ray band for the first time.    The morphology seen by \nustar\ agrees well with that seen 
by \xmm\ and \chandra\ below 10~keV.  At high energies \nustar\ clearly detects 
non-thermal emission up to $\sim20$~keV that extends along the eastern and northern rim of the supernova shell.   
The broadband images clearly demonstrate that X-ray emission from the North Spur and Eastern Limb results predominantly from non-thermal processes.  
We detect a break in the spatially integrated X-ray spectrum at  $\sim9$~keV that cannot be reproduced by current SED models, implying either a more complex electron 
injection spectrum or an additional process such as diffusion compared to what has been considered in previous work.  
We use spatially resolved maps to derive an energy-dependent cooling length scale, $L(E) \propto E^{m}$ with $m = -0.21 \pm 0.01$.  
We find this to be inconsistent with the model for the morphological evolution with energy described by \citet{kc84a}. 
This value, along with the observed steepening in power-law index between radio and X-ray, can be quantitatively explained as an energy-loss spectral 
break in the simple scaling model of \citet{rey2009}, assuming particle advection dominates over diffusion.  This interpretation requires a substantial 
departure from spherical magnetohydrodynamic (MHD), magnetic-flux-conserving outflow, most plausibly in the form of turbulent magnetic-field amplification.  

\end{abstract}

\section{Introduction}

A pulsar-wind nebula (PWN) is a bubble of relativistic particles and magnetic field inflated by a rotation-powered pulsar, emitting centrally peaked synchrotron radiation.
Young PWNe are frequently found inside shell supernova remnants (SNRs), where the relativistic wind of electron-positron pairs (and perhaps ions)
experiences a wind termination shock close to the pulsar due to the pressure of the SNR interior.  Beyond this wind shock, the relativistic fluid can radiate synchrotron 
emission from the radio to the X-ray band, and inverse-Compton (IC) emission at higher energies.  A PWN/SNR combination is often called a `composite' SNR. PWNe can outlive their  
SNR and interact directly with the interstellar medium.  See \citet{gs2006} for a review.

The detailed characterization of PWN spectra gives information on the particle energy distribution produced by the pulsar, and also on the nature of acceleration 
in relativistic shocks.  However, spectral structure present immediately downstream of the wind shock can be altered by propagation effects including diffusive transport 
and radiative losses, which may depend on 
the evolution of the entire PWN.  The ability to use PWNe as laboratories in which to study the behavior of relativistic pair plasmas depends largely on the extent to 
which these various effects can be disentangled.  Models studying PWNe with time-dependent, one-zone, homogeneous approximations, 
provide insights into the evolution of the spectrum across all energy bands of the nebula as a whole.
Broadband spectral-energy distributions (SEDs), coupled with spatially resolved spectroscopy, are required for this purpose.

Radially-dependant models also provide valuable insights by addressing detailed spatial and spectral structures of PWNe.
The classic work of \citet{kc84a}, 
which invokes a particular hydrodynamic model, 
predicts the spectral break between the optical and X-ray energy bands, as well as the 
behavior of size with photon energy for the Crab Nebula.  
Implicit in that calculation is a prediction for the radial dependence of the spectrum.  
\citet{kc84a} assume particle transport by pure advection in a spherical geometry; 
this situation predicts a roughly uniform nebular spectrum with radius until the (energy-dependent) nebular edge, where radiative energy losses sharply 
steepen the spectrum. This behavior  is generally inconsistent with observations \citep{rey2003, tc2012}, in particular in several objects a fairly uniform spectral 
steepening with radius is observed \citep[see][]{bb2001}.   

The cause of spectral structure is of interest beyond the specific PWN context; if its origin is in the physics of particle acceleration at relativistic shocks, 
the results have implications for other objects which share the same shock mechanisms. However, the structure may be due instead to transport and evolutionary effects 
in the PWN post-shock flow.    At the same time, spatially integrated spectra of PWNe are not well understood, especially in their most salient feature -- a large steepening 
of the spectrum between the flat radio emission and the considerably steeper spectra observed from IR through X-rays.  \citet{chev2005} 
documents differences between radio and X-ray spectral indices of $\Delta \equiv \alpha_x - \alpha_r \sim 0.7 - 1$ ($\rm{S}_{\nu}\propto\nu^{- \alpha}$) 
for seven out of the eight PWNe modeled, compared to radiative energy 
losses in homogeneous steady sources which can produce, at most, $\Delta = 0.5$.  If this break is not due solely to evolutionary effects and instead is due to at least,
in part, to intrinsic spectral structure in the particle distribution injected at the wind shock, 
it is implying something important and interesting about particle acceleration in pulsar winds.

\g21 was discovered in 1970 \citep{alt1970, wa1970} in the radio band, and first observed in the X-ray band in 1981 \citep{bs1981}.  It is a classic example of a Crab-like PWN:  
it has a filled, mostly symmetric spherical morphology centered on a pulsar.  
Measurements spanning 44 years show a well-characterized flat spectrum in the radio regime  \citep[e.g.][]{gd1970, bk1976, salter1989, Bandiera2001}, 
with recent observations reporting a spectral index of  $\alpha_{\text{r}}=0.0\pm0.1$ \citep{beit2011}.
The flux density at 1GHz is $6$~Jy \citep{cam2006}, and $^{13}$CO and H\textrm{I} analyses
determined the distance to the nebula to be $4.7\pm0.4$~kpc \citep{cam2006, tl2008}.  In this paper we adopt a distance of $5$~kpc.  

Observations with \chandra\ and \xmm\ find that the X-ray emission is dominated by a centrally peaked core 
that contains $\sim85\%$ of the 2 -- 8 keV flux.  The spectrum is described by a 
non-thermal power law with no evidence of line emission. The total unabsorbed
nebular flux is $F_{\rm x}(0.5-10 $~keV$)=9.35\times 10^{-11}$~erg~cm$^{-2}$~s$^{-1}$ \citep{slane2000, sh2001,war2001}.
Safi-Harb and coworkers detected spectral steepening in the nebula, indicative of synchrotron cooling. The innermost 
$0.5''$ radius region has a power law photon index of $\Gamma=1.43\pm0.02$ ($\alpha_{\rm x} \equiv \Gamma-1$) which softens 
to $\Gamma=2.13\pm0.06$ at a radius of $40''$ consistent with the edge of the nebula.  
This spectral softening is also visible in hardness ratio images \citep{msh2005}.  The photon index of $\Gamma \sim 2$ yields 
$\Delta \sim 1$.
 
The associated pulsar, PSR J1833-1034, was discovered in 2005, 35 years after the initial detection of the PWN.  PSR J1833-1034 has a 
$61.86$~ms period, $\dot{P}= 2.0\times10^{-13}$, $\tau_c=4.8$~kyr, and $\dot{E}=3.3\times10^{37}$\,erg\,s$^{-1}$ \citep{gup2005, cam2006}.
Despite the pulsar's high spin-down luminosity, pulsations have not been detected in the X-ray band.  
\chandra\ and \xmm\ observations detected a region of diffuse, uniform emission extending from the edge of the PWN at $40''$ out 
to $150''$ \citep[e.g.][]{slane2000}, with 
a non-thermal power law spectrum of $\Gamma\sim2.5$.  The flux is substantially dimmer than the PWN.  This symmetric emission 
was proposed to be an extension of the PWN itself \citep{war2001}, but the absence of coincident radio emission, and the recognition of the 
importance of dust scattering for the large column density toward G21.5-0.9, led \citet{bandbocc2004} and \citet{bocch2005} to model 
the smoothly distributed halo emission as due to dust scattering.  

Additionally, two regions of brightened emission were discovered near the outer edges of the halo, referred to as the North Spur and 
Eastern Limb \citep{war2001, sh2001} that could not be explained by dust scatter. These regions, located approximately $80''$ and $120''$, 
respectively, from the center of the nebula,
have week but detectable X-ray emission \citep{bocch2005, bocchetal2005, msh2010}.  Deep radio observations have not detected any
emission from either the halo or the Eastern Limb, while the North Spur can be clearly seen in a 1.4~GHz image \citep{beit2011}.
\citet{bocch2005}, and later \citet{msh2010}, found the spectrum of the North Spur comprises of a weak, low-temperature thermal component and a 
non-thermal continuum.  This knot of emission has been interpreted as the result of ejecta interacting with the H-envelope of the SN.    
The Eastern Limb, in contrast, was found to have only a non-thermal spectrum, and has spectral and morphological features that imply it is 
the limb-brightened region of the SN shell \citep{msh2010}.  
However, the low surface brightness of both the Eastern Limb and North Spur prevent their $<10$~keV continuua from being characterized more
specifically.

In this paper we present the first sub-arcminute X-ray images above 10~keV and corresponding X-ray spectroscopic studies of \g21.
$\S2$ discusses the \nustar\ observations.  $\S3$ presents our spectral analysis, while $\S4$ presents our image analysis
and $\S5$ details our search for the pulsar in the high-energy X-ray band.  Lastly, in $\S6$ we discuss the  spectral and spatial 
studies and how \nustar\ can shed light on the physics in a PWN and the natures of the North Spur and Eastern Limb.

\section{\nustar\ Observations}

\nustar\ observed \g21\ on four separate occasions for a total of 281~ks:  
2012 July 29 (ObsID 10002014003),  2012 July 30 (ObsID 10002014004), 2013 February 26 (ObsID 40001016002), and 2013 February 27 (ObsID 40001016003).  
The center of the remnant was located approximately $2'$ from the on-axis position for all observations so that the majority of the PWN was located on one of 
the four detector chips.  \nustar, which contains two co-aligned optic/detector focal plane modules (FPMA and FPMB), has a half-power diameter (HPD) of $58''$, an 
angular resolution of $18''$ (FWHM) over its $3-79$~keV X-ray energy range, and a characteristic FWHM energy resolution of 400~eV at 10~keV.  The field of 
view is $12'\times12'$ at $10$~keV as defined by the full width at half intensity.  
The \nustar\ nominal reconstructed coordinates are accurate to $8''$ ($90\%$ confidence level) \citep{har2013}.

Prior to our imaging and spectral analysis, we registered individual observations to J2000 coordinates using the central peak position of the PWN measured by \chandra: 
RA(J2000)$=18^{h}33^{m}33^{s}.54$, DEC(J2000)$=-10^{\circ}34'07''.6$ \citep{sh2001}. Note that the \chandra\ centroid position is $<1''$ 
offset from the radio pulsar position \citep{cam2006}.  The \nustar\ field of view for \g21\ is devoid of any visible point sources including the foreground star SS 397, which is located $\sim100''$ SW of
the center of the PWN. We determined the centroid position of the inner $30''$ radius region of the 
PWN in the $3-10$~keV band using the IDL routine {\bfseries gcntrd} so that the \nustar\ image is not contaminated by  
substructures such as the Eastern Limb and North Spur. 
The centroiding errors (90\% confidence level) are $\sim{\rm 3}''$ in both RA and DEC. We subsequently confirmed by fitting a circular 2-D Gaussian convolved with the 
\nustar\ PSF  (Section 4.1) that the centroid position of the PWN overlaps with the radio pulsar position within our uncertainty.

\section {SPECTROSCOPY}

\begin{deluxetable*}{lccccccc}[ht!]
\tabletypesize{\footnotesize}
\tablewidth{0pc}
\tablecaption{Spectral fitting of the \nustar\ \g21\ data}
\tablecolumns{8}
\tablehead
 {\colhead{ } & \multicolumn{2}{c}{$r\leq165''$}   & \multicolumn{2}{c}{$r\leq30''$} & \multicolumn{2}{c}{$r=30-60''$} & \colhead{$r=60-90''$} \\
 \colhead{Parameter} & \colhead{Power-law} & \colhead{Broken P.L.} & \colhead{Power-law} & \colhead{Broken P.L.} & \colhead{Power-law} & \colhead{Broken P.L.} & \colhead{Power-law} }
\startdata

$\Gamma_{1}$ & $2.039\pm0.011$& $1.996^{+0.013}_{-0.012}$  & $1.964^{+0.011}_{-0.012}$ & $1.852\pm0.011$ & $2.051\pm0.012$ & $1.98^{+0.02}_{-0.03}$ & $2.09\pm0.014$ \\
$\Gamma_{2}$ & $\cdots$ & $2.093^{+0.013}_{-0.012}$  & $\cdots$  & $2.099^{+0.019}_{-0.017}$ & $\cdots$  & $2.14^{+0.02}_{-0.03}$ & $\cdots$ \\
${E}_{\rm break}$  & $\cdots$  & $9.7^{+1.2}_{-1.4}$ & $\cdots$ & $9.0^{+0.6}_{-0.4}$ & $\cdots$ & $9.74\pm1.0$  & $\cdots$  \\
$F_{{\rm x}(2-8{\rm keV})}$ & $5.45\pm0.03$  & $5.27\pm0.08$  & $\cdots$  & $\cdots$ & $\cdots$ & $\cdots$ & $\cdots$ \\
$F_{{\rm x}(15-50{\rm keV})}$ & $5.47\pm0.03$  & $5.11\pm0.08$   & $\cdots$  & $\cdots$ & $\cdots$ & $\cdots$   & $\cdots$ \\ 
$\chi^2/\text{d.o.f}$ & $1.33 (5117)$  & $1.09 (5112)$  & $1.32 (3564)$ & 0.99 (3557) & $1.08(2257)$ & $0.960 (2255)$ & $0.942 (1924)$  \\

\enddata
\tablecomments{Spectral fitting of \g21\ with various extraction regions. The column header indicates the region from which the data was extracted.
$N_{\rm H}$ was frozen to $2.99\times10^{22} \text{cm}^{-2}$ for all fits.  Flux is listed in units of $10^{-11}{\rm ergs}$~${\rm s}^{-1} {\rm cm}^{-2}$.
The goodness of fit is evaluated by the reduced $\chi^2$.  The errors are 90\% confidence level.}
\label{tab:specfit}
\end{deluxetable*}

\begin{figure}[]
	\includegraphics[width=0.45\textwidth]{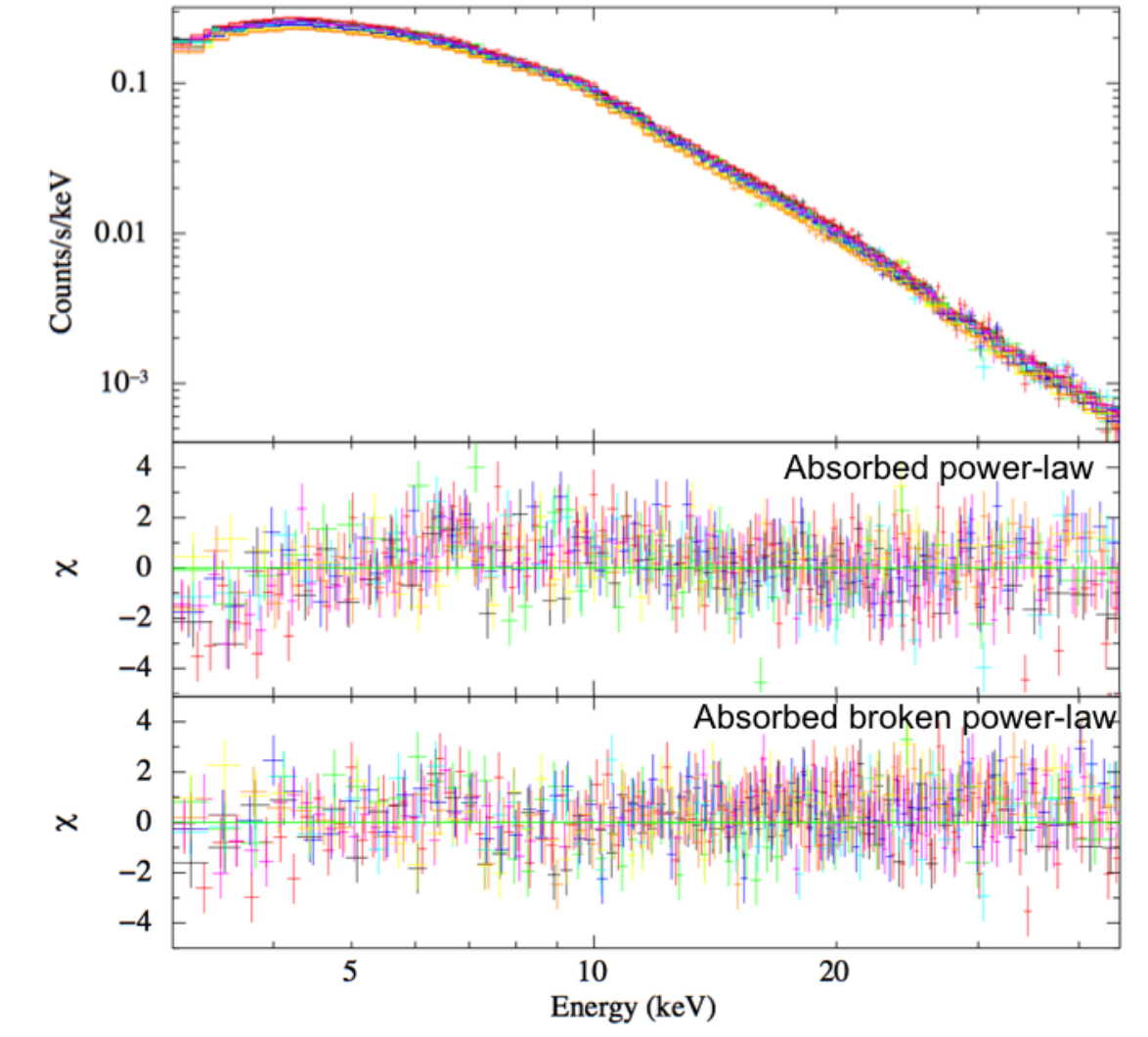}
\caption{ \nustar\ spectral fitting of \g21\ in the energy range $3-45$~keV from the entire PWN (extraction radius=$165''$).  Eight \nustar\ spectra 
and their residuals are shown.  The middle panel depicts the residuals fo for the absorbed power-law fit, while the lower panel depicts the absorbed broken power-law 
fit.
}
\label{fig:specFit}
\end{figure}

We performed \nustar\ spectral analysis integrated over the PWN region as well as 
spatially resolved spectroscopy which we present in subsequent sections. We applied the same analysis procedures, described below, for all
\nustar\ spectra. 

Prior to spectral fitting we generated \nustar\ response matrix (RMF) and effective area (ARF) files for an extended source using {\it nuproducts} (NuSTARDAS v1.1.1), 
then grouped the spectra to  $>20$ counts per bin using FTOOLS {\bfseries grppha} (HEAsoft 6.13).   
We fit the \nustar\ spectra using {\bfseries XSPEC} version 12.8.0 \citep{arnaud96}, with the atomic cross sections set to those from \citet{vern96}
and the abundances to \citet{wilms2000}. 
We fit the power-law model {\bfseries pegpwrlw} and the broken power-law model {\bfseries bknpower}. We multiplied the continuum models by the interstellar absorption model, 
{\bfseries Tbabs}, with $N_{\rm H}$ frozen to $2.99\times10^{22}$\,cm$^{-2}$ from \citet{tsuj2011}. 
We fit the spectra from each observation and each module jointly by 
linking all parameters except the continuum normalization, to take into account small calibration uncertainties between the two modules' flux normalization.

We generated background spectra using {\it Nulyses}, \nustar-specific software that accounts for detector background and cosmic X-ray background (CXB)
for a given extraction region of the source spectrum. 
{\it Nulyses} creates source-free background maps from the \nustar\ blank-sky survey data taken less than a month from the \g21\ observations.
This method was applied to all subsequent spectral analyses.
The conventional way of extracting background spectra from a nearby region is not applicable to our analysis since the PWN, broadened by 
the \nustar\ PSF, covers most 
of the detector chip and the detector background varies among the different chips.  
We fit the \nustar\ spectra in the $3-45$~keV band, above which the detector background dominates.

\subsection{\nustar\ spectroscopy of the entire PWN}

We extracted \nustar\ data using a $165''$ radius circular region centered at the pulsar \psr\ position.
This region represents the largest circle within the same detector chip (to ensure that \nustar\ detector background is the same and predictable), and it encloses 
over 93\% of the PWN photons. 

We first fit the \nustar\ spectra with a single absorbed power-law model in the $3-45$~keV energy band. 
We obtained a best-fit power-law index of $\Gamma=2.04\pm0.01$ and an X-ray flux of $F_{\rm 2-8~keV}=5.54\pm0.03$~erg~s$^{-1}$~cm$^{-2}$.  These results are consistent 
within their uncertainties to the parameters
reported by \citet{tsuj2011} ($\Gamma=2.05\pm0.04$, $F_{\rm 2-8~keV}=5.7\pm0.5$~erg~s$^{-1}$~cm$^{-2}$), 
who analyzed \g21\ for the purposes of cross-calibrating X-ray instruments operational at that time.
All uncertainties (90\% confidence level) quoted in the text and tables include both systematic and statistical errors. 
A more detailed comparison of \g21\ spectral fitting between \nustar\ and other X-ray telescopes as well as systematic error estimates will be addressed in a 
separate calibration paper. 

A single power-law fit is not satisfactory, with reduced $\chi^{2}=1.33$, with clear residuals evident between 5 and 10~keV (see the left panel 
of Figure \ref{fig:specFit}). 
We then fit a broken power-law model ({\bfseries bknpower}) to the \nustar\ spectra, yielding reduced $\chi^{2}=1.09$. Statistical tests using the
F-distribution ({\bfseries ftest} in XSPEC) show the broken power-law model is statistically favored over the single power-law model with high significance, so that
a spectral break is unambiguously required.  The best-fit power law indices were $\Gamma_{1}=1.996\pm0.013$ and $\Gamma_{2}=2.093\pm0.013$, with
the best-fit break energy of $E_{\rm break}=9.7\pm1.3$~keV.  

In order to investigate any instrumental effects that mimic a break around 9 keV, we analyzed \nustar\ observations of the known power-law source 3C273. 
We followed the same procedures used for \g21, and found no systematic residuals around 9 keV in the \nustar\ data. 
A spectral break in another PWN, the Crab Nebula, has recently been detected by \nustar\ at similar energies of $8-11$~keV \citep{madsen2014}.

Dust scattering can also mimic the low-energy spectral behavior 
observed in the \nustar\ spectra.  Recent analysis of two low-mass
X-ray binaries with similar absorption columns of 
$\sim3 \times 10^{22}$~cm$^{-2}$, GX5-1 and GX13+1, present radial
profiles of models of the dust scattering halos.  These profiles 
indicate that $0.1\%$ of the source photons at $2.5$~keV are enclosed 
between $50'' < r < 600''$ \citep{smith2002}.  
Including the energy-dependant halo intensity profile relationship
$I \propto E^{-2}$ \citep{ps1995, smith2008}, a radius of $r<160''$ encloses 
over $99\%$ of the source photons in both $3-5$~keV and $5-8$~keV.  
Similar analysis 
reveals that the $r<30''$ extraction region, presented in Section 3.2,
has a scattering loss of $\sim0.2\%$ and $\sim0.1\%$ of source photons in
$3-5$~keV and $5-8$~keV, respectively. 
Loss of photons by dust scattering therefore has a negligible effect, and cannot explain the 
spectral break in the \nustar\ spectrum.
Additionally, because our scattering estimates are based on point sources, we cannot conclusively rule out the possibility of a
very small effect of dust scattering on the radial behavior of the low-energy spectrum presented in Section 3.2.


An X-ray spectral break at $\sim12$~keV was first suggested by \citet{tt2011}.  They extrapolated the spectral fits from the hard and soft X-ray bands reported by \citet{tsuj2011},
and noted that a break at $\sim12$~keV was likely.  \nustar\ not only confirms the break, but provides the first measurement of the $\sim9$~keV break in 
broad-band 3 -- 45~keV spectra of \g21\  in a single X-ray telescope. 

\begin{figure}[t]
	\includegraphics[width=0.45\textwidth]{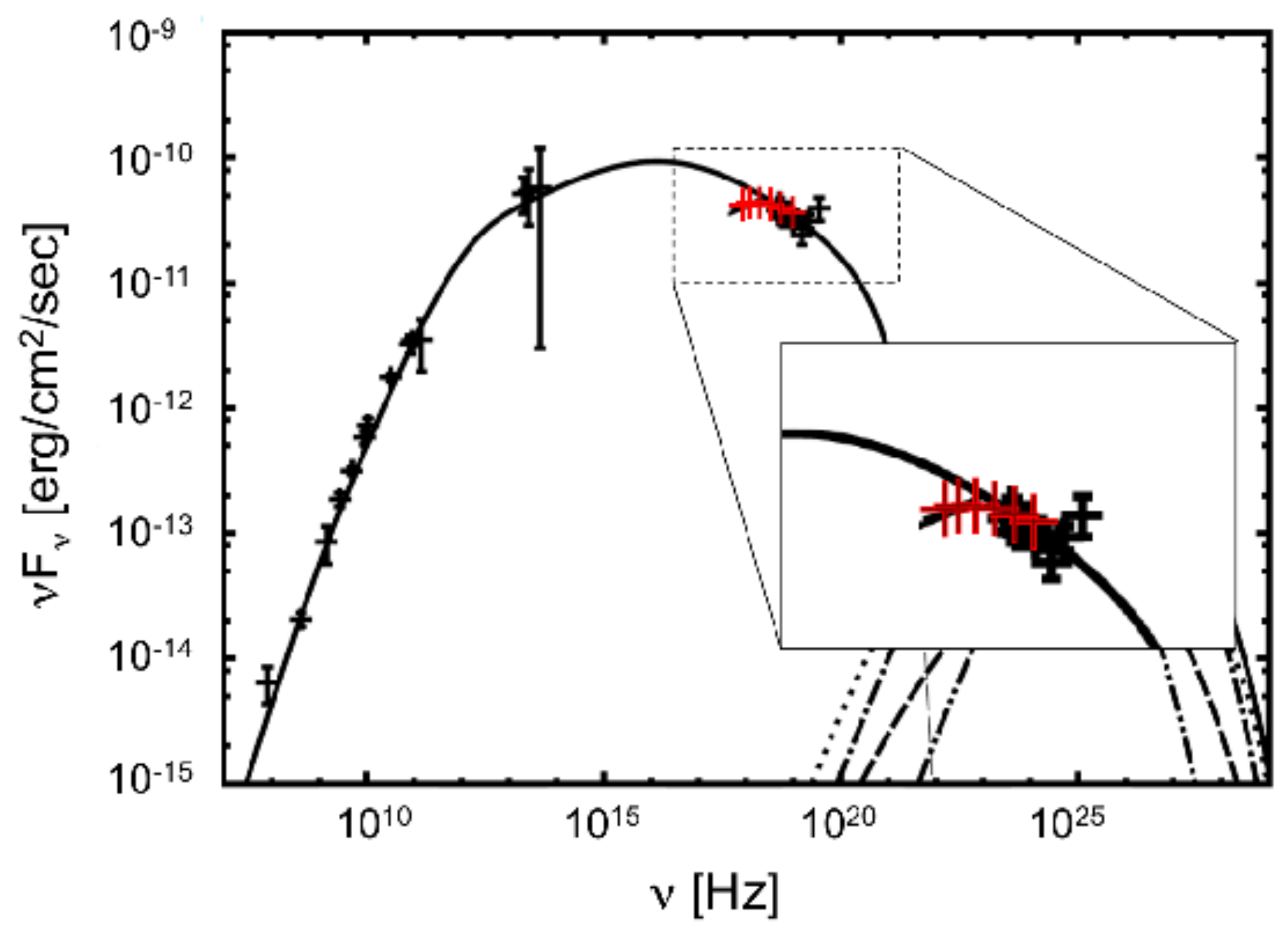}                                                                                      
\caption{ \nustar\ \g21\ data, represented by red crosses, overplotted on a graph containing data from previous observations, represented by black crosses, 
and a model prediction of the spectrum of \g21\ shown by the solid black line. From \citet{tt2011}.}
\label{fig:SED}
\end{figure}

Obtaining a well-calibrated X-ray spectrum from \g21\ is particularly valuable when creating PWN theoretical models.  Specifically, \citet{tt2011} presented 
synchrotron emission from radio to GeV energies by modeling the electron spectrum emanating from the 
pulsar of \g21.  They incorporated adiabatic and energy losses in a one-zone model.
Figure \ref{fig:SED} shows the \nustar\ best-fit broken power-law data overplotted on Figure 3 from \citet{tt2011}.  More details are presented in Section 6.1.

\subsection{Spatially resolved spectroscopy}

A number of young PWNe, including \g21, exhibit spectral softening from the center of the PWN outward. 
This is the classic signature of synchrotron burn-off. \citet{slane2000} and later \citet{sh2001} extracted \chandra\ spectra 
at various annuli of \g21\ and showed spectral softening from $\Gamma=1.43\pm0.02$ in the central $5''$ radius circle  
to $\Gamma=2.13\pm0.06$ in the outer annulus at a radius of $\sim40''$. 
\nustar\ has the ability to probe the synchrotron burn-off effects in the hard X-ray band above 10 keV. 

We extracted \nustar\ spectra from the central $30''$ region as well as from four nested annuli, each $30''$ in width. 
After experimenting with various annulus widths, we found that the annulus width of $30''$ is the best compromise between the \nustar\ angular resolution and the 
expected spatial variation from the \chandra\ results. 

\begin{figure}[t] 
	\includegraphics[width=0.45\textwidth]{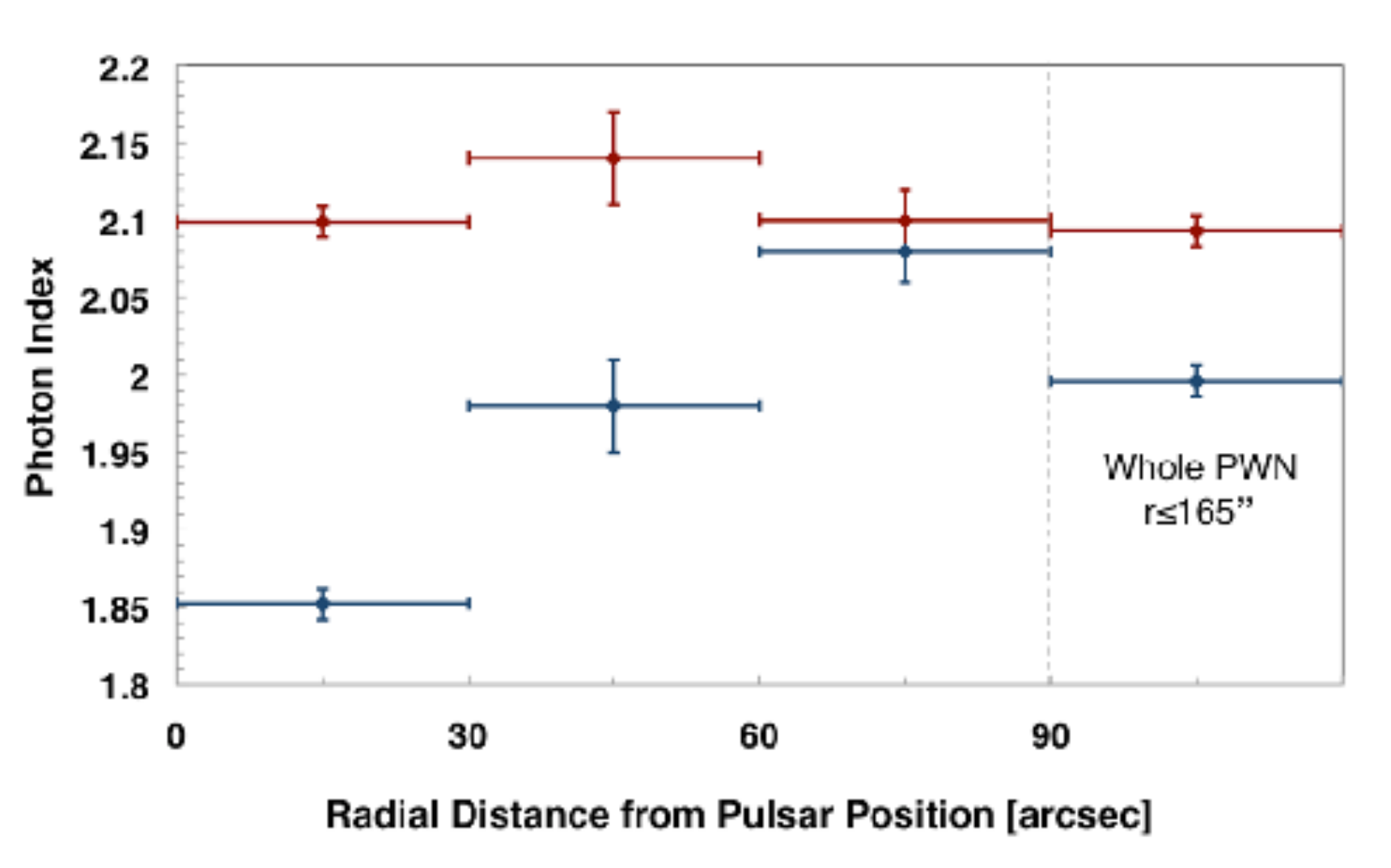}
\caption{ \nustar\ spectral photon indices as a function of extraction region for an absorbed {\bfseries bknpower} fit.  The radial positions are measured from the radio 
location of the pulsar, RA(J2000)$=18^{h}33^{m}33^{s}.54$, DEC(J2000)$=-10^{\circ}34'07''.6$.  The blue points illustrate the photon index $\Gamma_{1}$ below the spectral break of $\sim9$~keV.
The red points illustrate the photon index $\Gamma_{2}$ above the break.}
\label{fig:GGraph}
\end{figure}

The broad-band $3-45$~keV fit results from both the absorbed {\bfseries powerlaw} and {\bfseries bknpower} models are shown in Table \ref{tab:specfit}. 
The power-law index fit with a single power-law model increases  
from $\Gamma=1.97\pm0.01$ in the $30''$ inner region to $\Gamma=2.25\pm0.02$ in the outer $90-120''$ radius annulus, confirming the spectral softening discovered by \chandra. 
However, the residuals for the inner $30''$ and $30-60''$  regions also show a spectral break around 9~keV. 
We fit a broken power-law model, and the results are shown in Table \ref{tab:specfit} and plotted in Figure \ref{fig:GGraph}.
We conclude that the spectral break at 9~keV is detected with high significance in the inner $30''$ radius circular region and the 
$30''-60''$ annulus. Interestingly, as is shown in Figure \ref{fig:GGraph}, the second (high energy) power-law component shows no spatial variation. The spectral 
softening is observed only below 9~keV.

\section {IMAGING ANALYSIS}

In the  imaging analysis, we applied common procedures for image preparation.   Specifically, we
generated the following; 1) mosaic images from the four observations, 2) exposure maps with vignetting effects included 
and 3) an effective PSF that takes into account the
various off-axis positions, orientation angles, and exposure times of the observations. 
After proper source registration as described in Section 2, we selected photon events in different energy bands using {\bfseries dmcopy} (CIAO v.4.4).
We chose energy bands to ensure sufficient photon statistics in each image, as well as to minimize the effects of averaging the
energy-dependent vignetting function over the energy range.
We then summed all four \nustar\ raw images in each energy band to create mosaic images using {\bfseries  XIMAGE} (HEAsoft 6.13). 
For each observation, we generated an exposure map with the optics vignetting effects using {\it nuexpomap} (NuSTARDAS v 1.1.1) and summed those as well.
We then applied exposure-corrections to the raw mosaic images in {\bfseries XIMAGE } and generated energy-dependent flux images.  
Figure \ref{fig:RawProfiles} shows the $3-6$~keV flux image on the left, clearly showing the centrally peaked PWN, broadened by the wings of the 
\nustar\ PSF.  In order to investigate the radial profile and detect subtle features possibly buried in the raw images, we generated an effective PSF  used 
for image deconvolution and 2-D forward image fitting, as described in the subsequent sections.

We estimated the background level  using the {\it nulysis} software described in Section 3. First, we reproduced the
background levels in the three detector chips where the contribution of \g21\ is negligible. 
We found that the background for the \g21 observation is largely due to the stray-light CXB component below 20 keV and the internal
detector background above 20 keV.   We were able to 
reliably produce the background count rates on the detector chip containing the \g21\ image, and found that the source emission is dominant up to 30 keV. Hereafter 
we present imaging analysis below 30 keV.

\subsection{\g21\ PWN energy-dependent radius}

A measurement of the PWN radius in different energy bands is complementary to the radially dependent spectral analysis. 
The angular resolution of \nustar\ is comparable to the 
PWN size ($\sim80''$). Therefore it is essential to take into account the blurring effects by the PSF convolution. 

We adopted a forward-folding method to measure the size of the PWN using {\bfseries Sherpa},CIAO's modelling and fitting package \citep{frusc2006}. The use of {\bfseries Sherpa} 
allows us to use the goodness-of-fit test (C-statistics) to find the best-fit parameters. We convolved an assumed source model, a circular 2-D Gaussian profile, with the 
effective PSF described in Section 3, 
and fit the \nustar\ flux images. We also ran the {\bfseries conf} command in {\bfseries Sherpa} to determine 90$\%$ confidence intervals for all fit parameters.

\begin{figure}[]
\centering
		\includegraphics[width=0.45\textwidth]{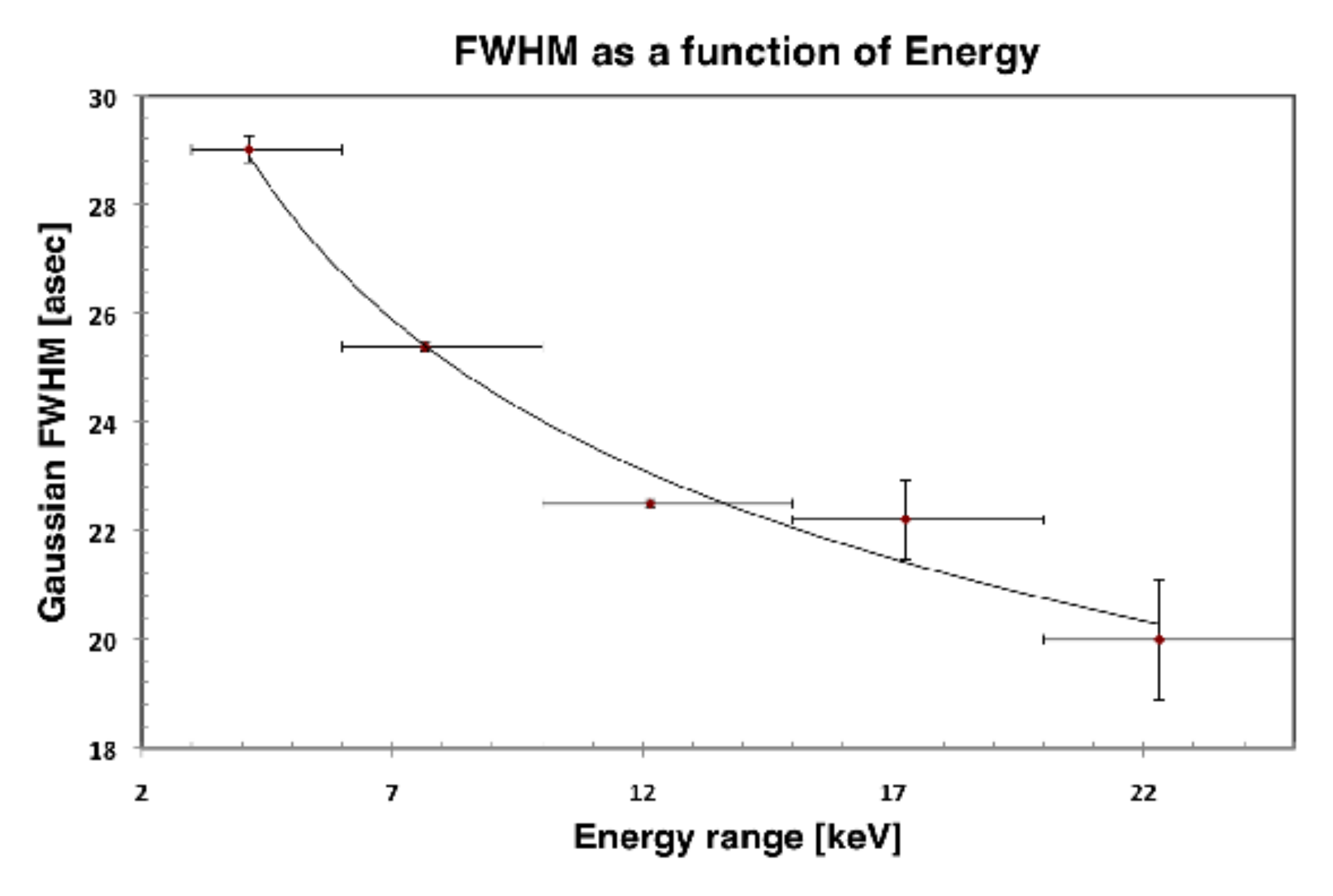}
\caption{2-D Gaussian FWHM radius of \g21\ as a function of energy.  The midpoint in each band is the mean energy, weighted by the \nustar\ flux.
The data is well fit to a power-law model $L(E)\propto E^{m}$, with $m=-0.21\pm0.01$. }
\label{fig:RvE}
\end{figure}

We confined our fitting range to within a radius of $160''$. In general, a circular 2-D Gaussian provides a good fit to the entire PWN, 
leaving only small residuals. The best-fit Gaussian FWHMs in five energy bands are 
plotted in Figure \ref{fig:RvE}. The trend of decreasing PWN size with energy is evident, confirming the synchrotron burn-off effect at energies above $10$~keV.  
In \S~6.2 we employ the FWHM radius to measure the synchrotron cooling length.
We fit the data with a power-law $L(E)\propto E^{m}$.  The best fit of $L(E)$ is shown in Figure \ref{fig:RvE} and yields an index of $m=-0.21\pm0.01$. These results 
are used in \S~6.2 to constrain physical conditions in the nebula.

\subsection{The North Spur and Eastern Limb }

\begin{figure*}[t]
  \centerline{ \hfill
	\includegraphics[width=0.4\textwidth]{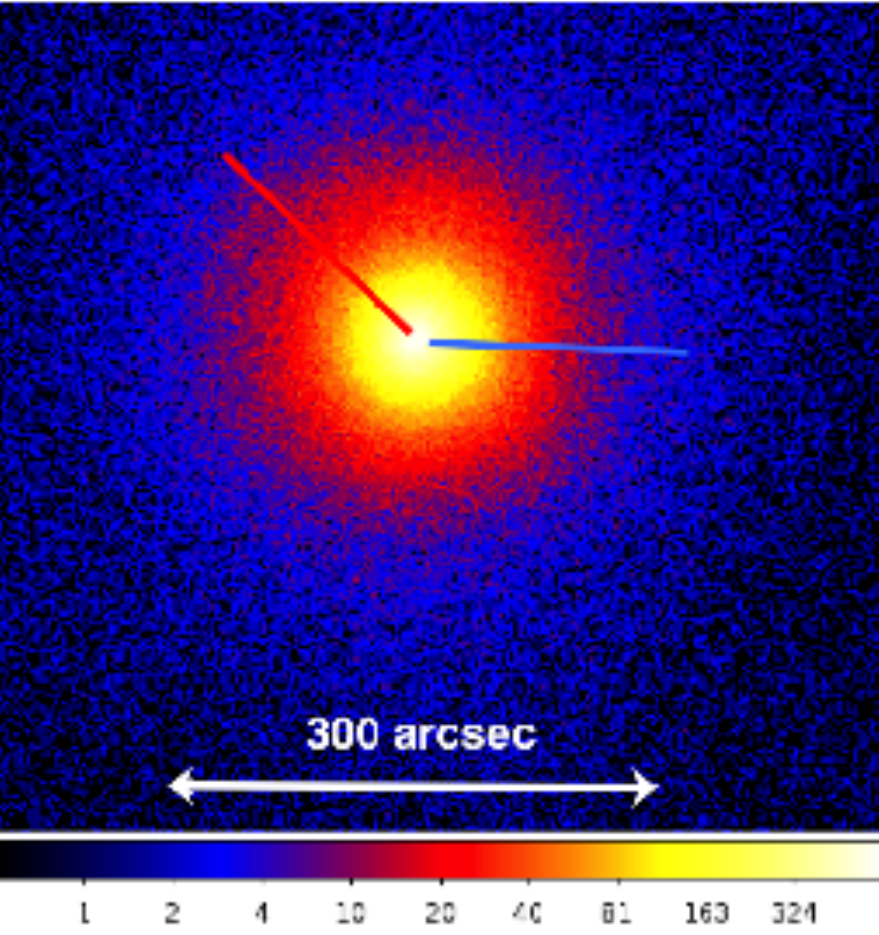} \hfill
	\includegraphics[width=0.5\textwidth]{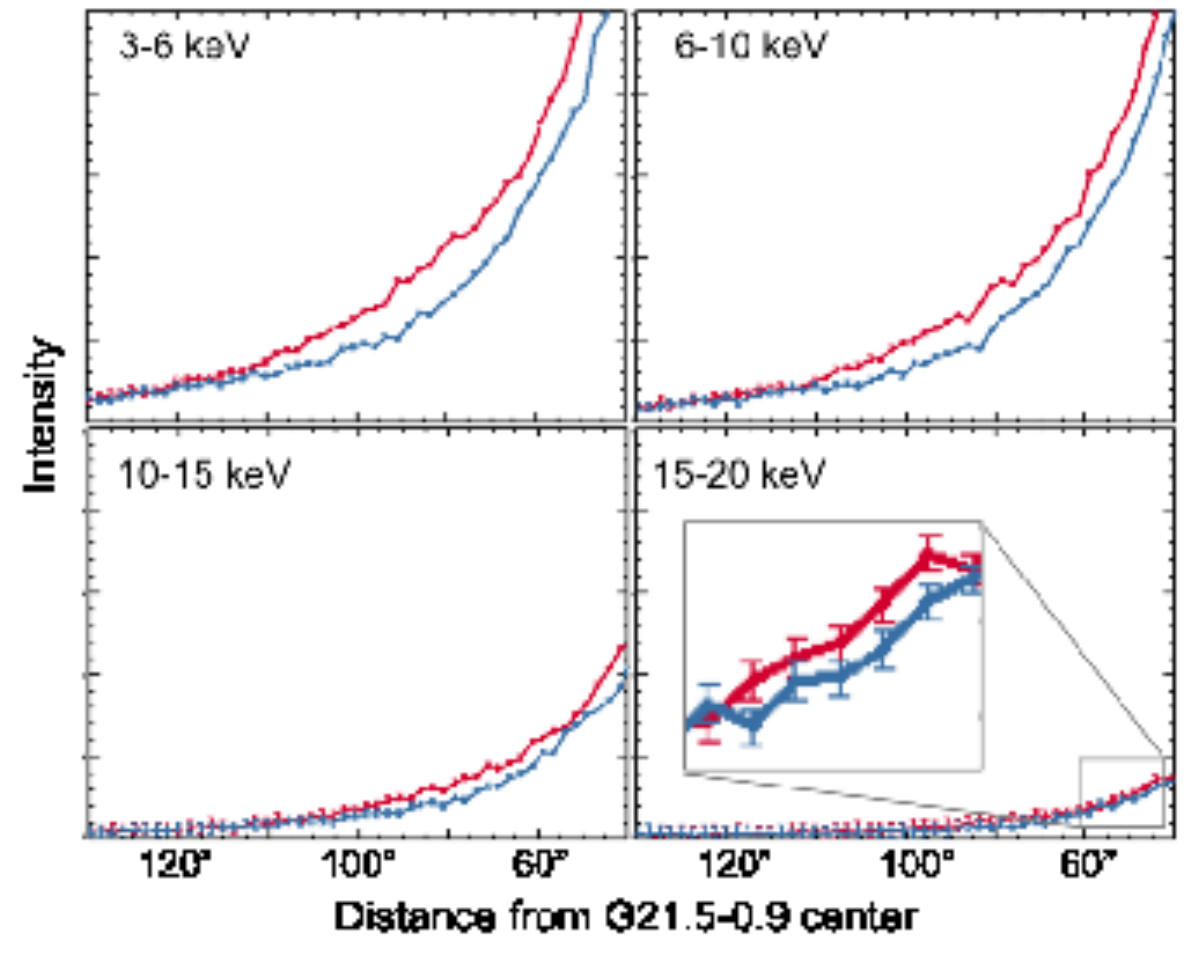} \hfill }
\caption{
Intensity profiles of \g21\ were obtained to confirm the existence of the Eastern Limb and North Spur in the raw \nustar\ images.  
{Left-- } $3-6$~keV \nustar\ mosaic image. Exposure-map vignetting corrections were applied, and FPMA and FPMB summed. 
The red and blue lines indicate the locations along which the profiles were obtained. 
{Right-- } Intensity profiles as a function of distance (in arcseconds) from the PWN center from the \nustar\ $3-6$~keV, $6-10$~keV, $10-15$~keV, $15-20$~keV image.  
The profiles show clear significant excess across the north eastern side of \g21, confirming the detection of excess emission. 
}
\label{fig:RawProfiles}
\end{figure*}

The Eastern Limb and North Spur are only partially resolved in the raw \nustar\ mosaics 
due to the small size of \g21\ compared to the \nustar\ PSF, combined with the low surface brightness of these features.  
To detect these features with high confidence, we applied two different methods. 
First, we confirmed the detection of the Eastern Limb and Northern Spur in the flux images in each energy band (Section 4.2.1). 
Second, we sharpened the \nustar\ flux images using the Lucy-Richardson deconvolution algorithm \citep{rich1972, lucy1974}.

\subsubsection{1-D profile analysis} 

In order to detect the Eastern Limb and North Spur in the raw \nustar\ image, we analyzed line intensity profiles of 6 image
pixels (a total of $\sim9''$) in width. We chose projection axis lines 
along different orientation angles around the center of the PWN. The left image in Figure~\ref{fig:RawProfiles} 
contains two lines, colored red and blue, that indicate the positions along which the flux profiles were taken.  The red line was chosen to fall 
along the regions of brightest Eastern Limb emission, while the blue line overlays a region of \g21\ that does not contain any emission from either 
the Eastern Limb or the North Spur.  The intensity profiles on the right of Figure~\ref{fig:RawProfiles} 
correspond to the lines drawn in the left of 
Figure \ref{fig:RawProfiles}.  The red profile shows clear and significant excess emission when compared to the blue profile, confirming that the
Eastern Limb is detected in the $3-6$~keV band.  This also holds for the $6-10$~keV and $10-15$~keV.  While the two profiles in the $15-20$~keV band 
are very similar, the lines are separate and distinct between $60''-40''$, when accounting for statistical uncertainty.  
With these intensity profiles we confirm the detection of the North Spur and Eastern Limb at energies as high as $20$~keV. 

We repeated this process with the North Spur, orienting the red line along the north/south axis.
Lastly, we also confirmed the detection of the Eastern Limb and North spur in residual maps from fitting the PWN, as mentioned in Section 4.1. 
The residual maps had areas of faint excess, indicating that there exists emission aside from the PWN.
However, the broad \nustar\ PSF prevents us from studying any further morphology of the faint emission.  We therefore turn to image deconvolution
in an attempt to remove the effects of the PSF.  

\subsubsection{Image deconvolution: method and verification}

We applied an iterative deconvolution technique to the \nustar\ images using {\bfseries arestore} (CIAO v4.4) and the effective PSFs described in Section 4.
The iterative image deconvolution can produce artificial features if the process is over-iterated and/or the background region is deconvolved. 
Great care was taken to ensure that the deconvolved image was both stable and reproducible.  We deconvolved the $3-6$~keV mosaiced \nustar\ image 
with several iterations, from 20 up to 200 in increments of 20.  Each of these images exhibited the same features 
of the North Spur and Eastern Limb that are characterized by areas of brightened emission to the North and Northeast of the PWN. 
Additionally, we confirmed that the features visible through deconvolution were not dependent on telescope rotation, detector module, or observation.
The data sets were first grouped by \nustar\ module, summing all the FPMA images before deconvolving, and likewise for all the FPMB images. 
The deconvolved images have identical features.  We also grouped the data by epoch, summing the 2012 and 2013 data separately.  
As before, the two deconvolved images are very similar.  

\begin{figure*}[!ht]       
\centering
	\includegraphics[width=0.98\textwidth]{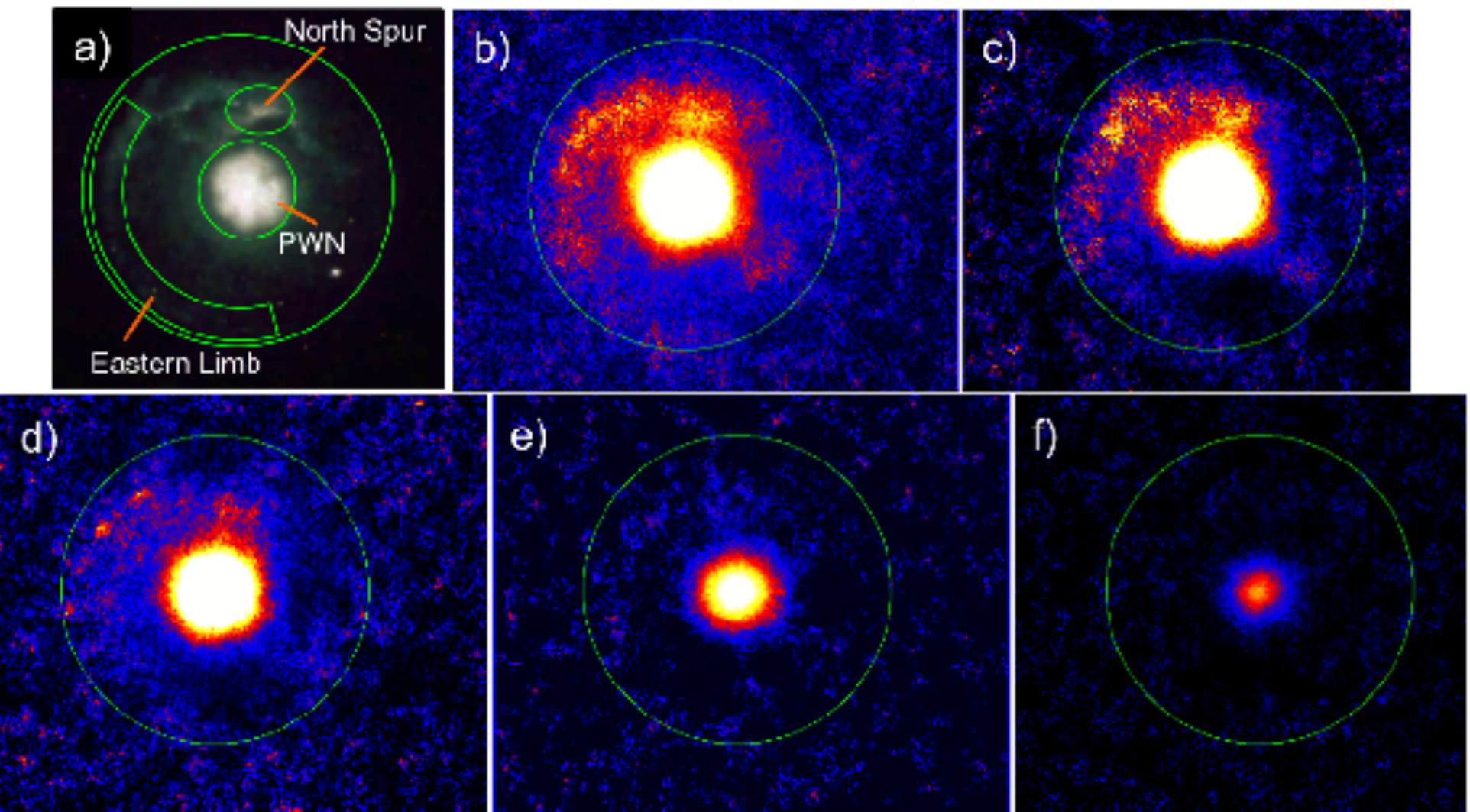}
\caption{ Deconvolved \nustar\ images at various energy bands: (b) 3-6~keV, (c) 6-10~keV, (d) 10-15~keV, (e) 15-20~keV, and (f) 20-25~keV. 
The images show the faint emission from the Eastern Limb and North Spur.  
The images are shown on a logarithmic scale, and colors were chosen to highlight the non-plerionic details.
Image (a) shows the \chandra\ $3-6$~keV image for comparison.  The green circle has a radius of $165''$.
}
\label{fig:Decon}
\end{figure*}

As a final verification of our deconvolution process, we compared the output images to the $3-6$~keV image from \chandra.  Figure \ref{fig:Decon} (a) shows the \chandra\
image at the top left corner.  One can see the Eastern Limb and North Spur, which is identified as `knot' in the image.   The North Spur is visible as the 
excess of emission $~100''$ north of the PWN.  The Eastern Limb is an arced feature that begins at the southern edge and wraps clockwise around to the northern edge,  
located approximately $\sim150''$ from the center of the plerion.  For reference, the outer green circle has a radius of $165''$. 
Figure \ref{fig:Decon} (b) shows the final deconvolved $3-6$~keV \nustar\ image, created with 20 iterations. The \nustar\ features correspond well to those seen by \chandra, allowing us
to confidently extend this deconvolution technique to higher energies.

\subsubsection{Image deconvolution: analysis}

Once we established that the \nustar\ results reproduce the \chandra\ images 
stably for a range of Lucy-Richardson iterations, we applied the same deconvolution to the higher energy images. 

Figures \ref{fig:Decon} (b-f) show the \nustar\ deconvolved images at $3-6$~keV, $6-10$~keV, $10-15$~keV, $15-20$~keV, and $20-25$~keV, respectively.  
One can visually identify the Eastern Limb and North Spur at energies as high as $15$~keV, as seen in Figure \ref{fig:Decon} (d).  
This is not surprising, since all of the 
previously reported spectral fitting of these two regions have required a 
non-thermal component.  This is, however, the first direct measurement showing that these features have emission above 10 keV. 

\begin{figure*}[]
\centering 
	\includegraphics[width=0.98\textwidth]{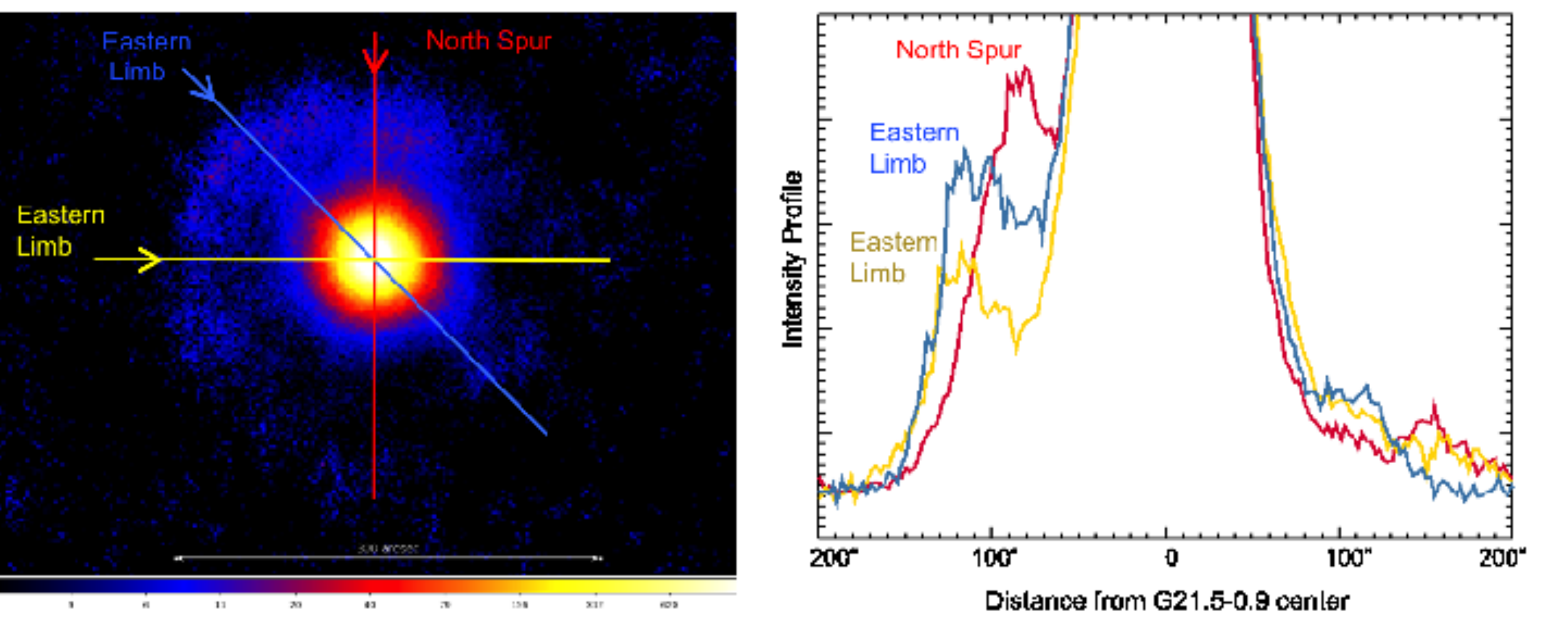}
\caption{{Left--} \nustar\ $3-6$~keV deconvolved image. Intensity profiles were obtained along three lines, shown in yellow, blue, and red, oriented at 
PA=$0^{\circ}, 45^{\circ}, and 90^{\circ}$, respectively.
{Right--} Intensity profiles obtained from the deconvolved \nustar\ images.  The profiles correspond to the lines of the same color, as shown on the left.  
Scaling was chosen to highlight the Eastern Limb and North Spur.}
\label{fig:DeconProf}
\end{figure*}

At energies above $15$~keV the Limb and Spur become very faint.  One can still see emission along the eastern edge of the Limb in Figure \ref{fig:Decon} (e).
We can, however, verify the existence of these features above $15$~keV 
by producing intensity profiles of the deconvolved images at various azimuthal angles.  Applying the same 1-D profile method as in Section 4.2.1, we  
obtained profiles along three different radial lines, each bisecting \g21\ at angles in increments of $45^{\circ}$ as shown
in the left image of Figure \ref{fig:DeconProf}.   The arrows overlaid on the lines indicate the direction along which the profiles were
obtained. 

\begin{figure*}[]       
\centering
	\includegraphics[width=0.98\textwidth]{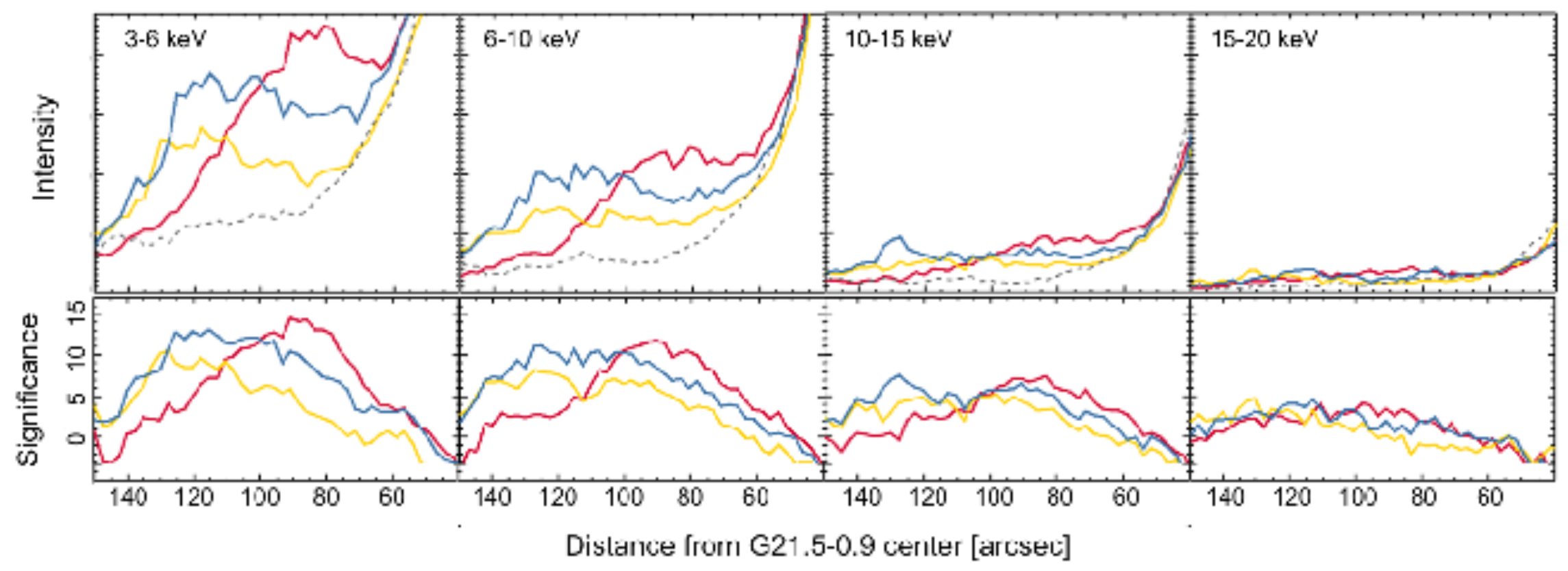}
\caption{ Intensity ine profiles and significance of \nustar\ deconvolved images.  
Excess emission indicating the detection of the Eastern Limb and North Spur is visible in all energy bands. The colors correspond to the same
angles as indicated in Figure \ref{fig:DeconProf}. The grey dashed line is the background emission, taken from the East side of the PWN, where 
there is no emission from either the Eastern Limb or the North Spur.}
\label{fig:DeconProfAll}
\end{figure*}

The intensity profiles themselves for the $3-6$~keV image are shown on the right in Figure \ref{fig:DeconProf}.  Each profile corresponds to the line of 
matching color overlaid on the decononvolved image on the left.  One can clearly see a sharp increase corresponding to the Eastern Limb in the blue and yellow
profiles, while the red excess indicates the existence of the North Spur.  Similarly, the Eastern Limb and North Spur are clearly visible in the profiles 
taken from the images up to energies of $15$~keV (see Figure \ref{fig:DeconProfAll}).   The intensity profiles from the $15-20$~keV image also show statistically 
significant photon excess from the Limb and Spur, confirming the existence of these features to energies as high as $20$~keV.

\section {TIMING SEARCH}

The flux in the $0.2-10$ keV band from the wind nebula generated by PSR~J1833$-$1034 completely dominates the pulsar
itself. The pulsar is barely resolved from the PWN at arcsecond resolution by \chandra\ and completely swamped by the PWN emission for other
X-ray telescopes.  All previous searches for X-ray pulsation have been unsuccessful despite extensive X-ray data sets collected with
sufficient timing resolution. Given the evidence of synchrotron burn-off above $\sim 10$~keV presented in Section 4.4, the extended energy band
of \nustar\ presents a new opportunity to further isolate the pulsar signal from the PWN.

To search for the signal from PSR~J1833$-$1034, we initially selected photons in the energy range $10<E<70$~keV from a small source extraction
aperture ($r<10^{\prime\prime}$) (see Figure \ref{fig:RvE}). We searched these photons for significant power around the expected period using the two
derivative radio timing ephemeris presented in \citet[][Epoch 2009]{abdo2010} extrapolated to the \nustar\ observation epoch. This
solution is preferred over the slightly updated five derivative model of \citet{Ack2011} whose extrapolated behavior is not predictive.
Lacking a coeval ephemeris, it is not possible to maintain phase unambiguously over the 206 day gap between the two \nustar\
observations. Instead, we separately search data collected in 2012 and 2013 which span $5.8$~d and $2.65$~d, respectively.

Taking into account the increased uncertainty in the timing parameters for the extrapolated ephemeris, we searched for a significant signal
over a frequency range of $\pm3 \times \sigma_{f}$, where $\sigma_{f}$ is the uncertainty in the frequency measurement,
oversampled by three times the Fourier resolution. We evaluate the power at each frequency
using the $Z^2_n$ test statistic for $n=1,2,3,5$, to be sensitive to both broad and narrow pulse profiles, possibly single or double
peaked.  The most significant signal in this search range was $Z^2_5 = 19.34$ and $Z^2_3 = 21.05$, using 6.4 kcts and 13.2 kcts, 
respectively, for the 2012 and 2013
observations. This corresponds to a significance of $0.72$ and $0.043$ after taking into account the number of search trials, $20$ and $24$,
for the two observations. We repeated our search for a additional combination of energy ranges $10<E<20$~keV, $20<E<79$~keV and aperture
size with radius $<20^{\prime\prime}$ but find no signal with a greater significance than the initial $2\sigma$ result. We conclude that no
pulsed X-ray signal in detected from PSR~J1833$-$1034 in the optimal \nustar\ band and place an upper limit at the $99.73\%$ ($3\sigma$)
confidence level on a sinusoidal signal pulse fraction of $f_p \approx 4.2\%$ and $6.1\%$ (including unknown PWN emission) for the two observations,
respectively.  

\nustar\ is not able to independently measure the spectrum of the pulsar.  However, \citet{msh2010} were able to isolate and fit the spectrum of
a $2''$ region located at the cite of the radio pulsar.  Using their best-fit non-thermal spectrum we can approximate the contributions of the pulsar
and PWN in the \nustar\ region.  This increases the pulse fraction to $f_p \sim 19.2\%$ and $27.9\%$, respectively.

\section {DISCUSSION}

\subsection{PWN spectral break and softening}

\chandra\ observations of \g21\ show spectral softening over the PWN, with the photon spectral index ranging from $\Gamma\sim1.4$ at the inner (radius $<5''$) region to 
$\Gamma\sim2.1$ at the outer $35''-40''$ radius annulus. 
\nustar\ observations confirm this by showing spectral softening below $9$~keV, as shown in Table \ref{tab:specfit}.
The broad \nustar\ PSF causes some mixing of the spectra in different annuli, and as a result the variation of $\Gamma$ with radius
obtained by \nustar\ is less pronounced than that seen by \chandra.  
\nustar\ finds a spectral break at $9.7$~keV in the integrated PWN emission.  The spectral break is also observed in the inner regions with radius $<30''$ and 
in an annulus from $35''-40''$.  The break is statistically significant at radii $>40''$.  Above $10$~keV the photon spectral index remains constant.

There are several possible origins for the spectral break seen by \nustar.  
One possibility is that the \nustar\ PSF mixes the radially dependent power-law indices seen by \chandra\  \citep{slane2000, sh2001}, and softens the spectrum 
in such a way as to cause a sharp break. However, simulations that fold \chandra\ maps through the \nustar\ response indicate that this is not the case.
The effects of the large ISM extinction \citep{tsuj2011} again would not cause such a sharp, defined energy break.  
Similarly, the loss effect of dust scattering at lower energies is
negligible (below $2\%$)in the \nustar\ band because of the relatively
low column absorption.  
Contributions from the pulsar are also not likely to be responsible. Most pulsars have a harder spectrum than the PWN they power.  While this could in fact 
cause a sharp transition between photon indices, it would cause spectral hardening with energy, not the softening seen by \nustar.  
Finally, spectral breaks are often attributed to synchrotron cooling, as proposed by \citet{tt2011}.  
However, the break energy of 9~keV would require an unreasonably low magnetic 
field strength of $\sim6$~$\mu{\rm G}$, compared to the $\sim300$~$\mu$G derived from equipartition arguments \citep{cam2006}.  
The prominent spectral steepening between radio and X-rays, if interpreted with simple cooling models, yields magnetic field strengths ranging from 
$25\mu {\rm G}$ \citep{dj2008} to $64\mu {\rm G}$ \citep{tt2011}.

It is likely that the spectral break results from physical effects, either due to a break in the injected electron energy spectrum or due to energy losses
due to particle transport in the PWN.   Pulsars emit pairs of relativistic electrons and positrons, which are accelerated 
at a termination shock near the pulsar itself.  Downstream of the termination shock the accelerated electrons interact with the magnetic field, also produced by the pulsar, and 
subsequently emit synchrotron radiation from the radio through gamma energy bands.   The injection spectrum can therefore shape and influence the spectrum of the synchrotron radiation, 
as noted by \citet{tt2011}.  These authors propagate a broken power-law electron spectrum through a time-dependent model that includes energy losses due to synchrotron  
radiation, inverse Compton scattering, and adiabatic cooling.
\nustar\ provides confirmation that the relatively simple injection spectrum used in \citet{tt2011} is not adequate to fit the observed X-ray data, as seen in Figure \ref{fig:SED}.
The model, noted as the solid black line, has a steep negative slope in the X-ray band and does not fit the X-ray spectra obtained from \chandra, \XMM, \integral, and now \nustar.  

It is therefore reasonable to explore whether a more complex model
can explain the 9~keV spectral break. \citet{vorster2013} extend the models with the addition of diffusive losses as well as 
a broken injection spectrum with a discontinuity at the break energy. However, both the aforementioned SED models do not include the spatial dependence of parameters 
such as the magnetic field of the PWN, which provides additional complexity and can perhaps better fit the X-ray data.  We explore these effects in
the following sections.

\subsection{Physical conditions inferred from cooling scale length measurements}

Particle transport in PWNe has long been a matter of discussion and study. Unfortunately, few celestial objects are both bright enough and 
large enough to allow distinguishing radially-dependant features to fit to the various existing models.  The Crab Nebula, \g21\, 
and 3C 58 are such PWNe.  They have been frequently observed and analyzed, and have provided valuable insights 
into the physics of magnetohydrodynamic (MHD) flows and relativistic
shocks that govern the appearance of PWNe \citep[e.g.][]{kc84a, kl2004, chev2005, msh2010}.
Extending the energy range of analysis will allow for more detailed probing of radially-dependant physics.
This information is important both to understand PWNe and to isolate the properties of the relativistic shock at the PWN inner edge from
the subsequent spectral evolution downstream.  We shall ask: can the observable parameters of \g21 from radio to X-ray be reproduced assuming the shock
injects only a single structureless power-law particle spectrum?

One such parameter used as a spectral fingerprint of various models is the radial dependency of the power-law photon index.
Numerous observations have confirmed softening of the \g21\ PWN spectrum with increasing projected radius \citep{sh2001, war2001, msh2005, slane2000, msh2010}.  
This is shown at higher energies for the \nustar\ observations in Figure \ref{fig:RvE}. The softening is associated with synchrotron cooling of the electrons, 
and the corresponding decrease in the maximum emitted X-ray energy in the bulk velocity flow downstream of the termination shock. 

An alternative approach is to characterize the variation of source size with photon energy using a characterization of
the `cooling length' $L(E)$, such as the source FWHM.  Since only the energy-dependence 
of this length is important for modeling the spectral steepening, its precise definition is not important. This scale depends only on mapping the total number of
photons as a function of radius and energy to determine the scale length $L(E)$.  This is more straightforward than spectral modeling because counting statistics are almost 
always limited, and length scale measurements do not require determining a spectrum at each radius.  
We therefore use the cooling length scale, with FWHM as its surrogate, as the fundamental diagnostic for extracting information about physical conditions. 

We shall also make use of the observed steepening from radio to X-rays of \g21\, through the parameter $\Delta \equiv \alpha_x - \alpha_r$.  
Most discussions of PWN physics \citep[e.g.][]{chev2005} simply take this as an intrinsic property, but we shall attempt to 
explain it through evolutionary effects.

There exist two prominent mechanisms that have been invoked to explain PWNe particle transport:  advection and diffusion. 
Beginning with \citet{wilson1972} and \citet{gratton1972}, diffusion has long been investigated as a cause for the characteristics of a PWN.  However, the early 
models of particles propagating outwards from a central source purely by diffusion, applied exclusively to the Crab Nebula,
have been unable to account for the detailed X-ray properties of the 
Crab, such as its change of size with frequency \citep{ku1976}. 
Pure advection models have also been proposed, such as \citet{rg1974}.  The canonical theory involving pure advection as the method of particle transport was 
presented by \citet[hereafter KC84]{kc84a}.  When KC84 was used to predict the radial behavior of the spectrum, the resultant
spectral photon index has little to no variation from the center of the nebula outwards, and begins to vary only towards
the PWN periphery  \citep[hereafter TC12, Figure 2]{tc2012}.  This also does not match the observed behavior of a slowly-steepening X-ray spectrum \citet{msh2005}.

TC12 provide a nuanced approach to particle transport by diffusion and present two updated models.
They claim that the magnetic field is not predominately toroidal far from the termination shock, as is often assumed in PWN theory, 
but has a more complex geometry with cross-field transport best described by diffusion.  
Their first model consequently incorporates pure diffusion only, with both the magnetic field and diffusion coefficient constant with radius, 
and synchrotron emission as the only loss of energy.
TC12 argue that such a model better explains the radial dependence of $\Gamma$ as seen in the Crab, 3C58, and \g21, with 
proper adjustment of the diffusion coefficient.   Involving complexities such as an energy-dependent diffusion coefficient
might be more physically reasonable, but the data are not sufficiently constraining to distinguish these cases from a simple diffusion model.  

While the pure diffusion model of TC12 appears to describe observations of the Crab and 3C 58 relatively well, we argue that it is less appropriate for \g21.
It does provide a good description of $\Gamma(r)$ for the Crab and 3C 58 ($\chi^{2}\sim1$), however the fit of this model to the \g21\ $>10$~keV data 
is poor ($\chi^{2}\sim3$), as seen in Figure 6 in TC12. 
In addition, the ratio of advective to diffusive timescales that determines what process dominates is radially dependant.  Using $v \propto 1/r^2$ downstream 
of the PWN termination shock (KC84), and the know relationships $t_{\rm adv} \propto r/v$ and $t_{\rm diff} \propto r$, we obtain $t_{\rm adv}/t_{\rm diff} \propto r$.  
\g21\ is very compact, with a size more than two
times smaller than that of the other two PWNe, and thus advection is likely to dominate.  

Finally, the advective
model provides a good fit to the energy dependent cooling length scale (see below).  
TC12 presented the radially-dependant spectral index $\Gamma(r)$ rather than using the cooling length scale $L(E)$ as 
characteristic of their model, which makes fitting the model to higher energy \nustar\ data too difficult.

The second model presented by TC12 involves a Monte Carlo simulation that includes both diffusion and advection transport close to and farther away from the pulsar, respectively.  
This allows a more complex treatment than previous analytical models \citep{massaro1985}.  While likely a more appropriate approach for \g21, TC12 only applied this simulation 
to the Crab and 3C 58.

KC84 provide a complete advective solution for a steady, spherically symmetric wind terminated by a magnetohydrodynamic (MHD) shock.
Although  KC84 is routinely applied to determine quantities such as the mean downstream magnetic field, its range of applicability is in fact more narrow. 
KC84 represents an idealized theory,  suited to the case of constant injection of electrons in a spherical geometry, 
transported outward by pure advection in an ideal MHD flow with an ordered, toroidal magnetic field.  
The model does not attempt to reproduce the Crab spectrum from radio to X-rays, but only
the optical to X-ray portion, with an injection spectral index of
optically emitting electrons of $\alpha_o = 0.6$.  The predicted value
of steepening of $\alpha_x - \alpha_o$ of 0.51 is roughly appropriate
(so their cooling break is at UV wavelengths).  This value is
fortuitously close to the value of 0.5 for a stationary, homogeneous
source.  Thus this model also cannot reproduce radio-to-X-ray SEDs of
many other PWNe, including \g21, which have a larger $\Delta$.  As mentioned above, the $\Gamma(r)$ relationship
predicted by the KC84 model is flatter in the PWN interior and
softens towards the edges, as shown by \citet{rey2003, tc2012}, while generally a gradual variation in $\Gamma(r)$ is observed. 
This has motivated several generalizations of KC84.  

The values we obtain from the \nustar\ data for $L(E)$ and $\alpha_{\text{x}}$ are clearly inconsistent with the model presented in KC84. 
The gradients implied by the assumptions of KC84, predict $L(E) \propto E^{-1/9}$, independent of spectral index, 
and $\Delta = (4 + \alpha)/9$ for the physically important inner flow region.  
If we attempt to apply the KC84 formalism to
describe the radio-to-X-ray spectrum of G21.5-0.9 with $\alpha_r =
0.0$ and $\alpha_x = 0.9$, we fail on both counts.
First, we find $m = -0.21 \pm 0.01$ ($\S4.1$). Second, we find the observed spectral indices in the 
radio and X-ray bands produce $\Delta=\alpha_{\text{x}}-\alpha_{\text{r}}=0.9\pm0.1$, instead of 5/9 from a KC model using the 
injected spectral index of $\alpha=1$.

\citet[hereafter R09]{rey2009} noted that there are a number of physical effects not accounted for in KC84 that could produce a steeper spectral break
than  $\Delta=0.5$.   For example, the magnetic field may have a significant radial or turbulent component, or
it may not satisfy mass
conservation due to cloud evaporation, or it may not have magnetic flux conservation (e.g. due to magnetic reconnection or turbulent amplification).  
R09 constructed a simple model that includes these effects.
The model involves generating simple scaling relations for the downstream magnetic field $B$, fluid velocity field $v$, and 
fluid density $\rho$ in terms of the dimensionless length scale 
$L = r/r_{\circ}$ where $r_{\circ}$ is the inner injection radius. The non-spherical 
geometrical effects could also be parameterized in terms of jet width $w=w_{\circ}L^{\epsilon}$ where $\epsilon=1$ corresponds to a 
conical jet or a section of spherical outflow; a confined jet has $\epsilon < 1$, while a flaring jet would have $\epsilon > 1$.  
By assuming $B$, $v$ and $\rho$ all vary as power-laws in $L$ with indices $m_{b}$, $m_{v}$ and $m_{\rho}$, R09 obtains 
a series of general consistency relations that these indices must satisfy with each other and with observable parameters.
These variables of $m$, $\alpha$ and $\Delta$, the energy scaling of the 
cooling length, particle injection index and spectral index break, respectively, provide constraints on the allowed values of
$\epsilon$, $m_b$, $m_v$, and $m_\rho$.

The results of R09 can be rewritten in terms of the functional form of $m$, and assuming $\alpha_r = 0$ as observed for G21.5-0.9:  
$$\Delta = (-m)(1 + 2\epsilon + m_\rho +  m_b)/\epsilon$$
independent of $m_v$.  Since we observe $\Delta = 0.9$ and $m = -0.21$, this gives $1 + m_\rho + m_b = 2.29\epsilon$. 
The possible solutions are thus very restrictive for our measured value of $\Delta$ and $m$.
Additionally, if the flow conserves mass (disallowing, for instance, mass loading by evaporation of thermal
material), then the density index $m_\rho$ is linked to the velocity index $m_v$.  In this case, some solutions are
unphysical, such as those with $m_{v}>0$, corresponding to an accelerating downstream fluid. With some judicious 
rejection of such solutions we can draw some interesting general conclusions, based on the possible values of the observables, which may guide further 
investigations.  

For a conical or spherical flow, $\epsilon = 1$, so either density or magnetic-field
strength, or both, must {\sl rise} with radius (since either
$m_\rho$ or $m_b$ is positive, or both).  Mass conservation links $m_\rho$ to $m_v$, but if that assumption is abandoned, there is no
relation and no constraint on the velocity profile -- only the density profile matters for observable quantities.  A steeply decelerating
flow can produce $m_\rho > 0$ with mass conservation, but mass loading can also do this.  Similarly, $m_b > 0$ can be a reasonable outcome of
flux non-conservation through processes such as reconnection.  Our observations require one or both of these effects: addition of mass to
the flow through some kind of evaporation, and increase in magnetic-field strength beyond flux freezing.

A strongly confined jet ($\epsilon < 1$) can relax some constraints; for $\epsilon = 0.3$, we only require $m_\rho + m_b = -0.32$.  However,
even here either mass or flux nonconservation is necessary. One can construct constant density solutions 
($m_{\rho}=0$)
but such a geometry is disfavored due to the high symmetry of \g21\ as evidenced in both the broad axial symmetry
observed in the radio \citep{furst1998} and in the soft X-ray band \citep{sh2001}.

While \citet{rey2009} expands on the treatment of KC84, both
fail to reproduce the steadily steepening spectrum with radius shown by G21.5--0.9 and other PWNe.  This shortcoming is characteristic 
of models in which particles are transported outwardly purely by advection, so that all particles at a given radius have similar ages. 
To produce steady spectral steepening probably 
requires a mixture of particles of different ages at each radius. 
This could be caused by more complex fluid flow such as the back flows found in simulations by 
\citet{kl2004}, or by particle diffusion.

We conclude that a model describing both the radio-to-X-ray spectrum of \g21\, and the size shrinkage with X-ray energy we observe can
be accommodated in a pure advection model requiring the injection of only a straight power-law spectrum of electrons, 
$N(E) \propto E^{-1}$.  However, as with all pure advection models, the gradual
rather than sudden steepening of the spectrum with radius is not reproduced.  The viability of this explanation for the observed
properties of \g21\ will depend on whether the addition of
diffusion can reproduce the gradual steepening while preserving the successes of the advection model.

\subsection{The North Spur}

\nustar\ has detected, for the first time, the North Spur and Eastern Limb above 10~keV.  Three main theories have been proposed to explain the 
nature of these features: they are extensions of the PWN itself, they are limb-brightened shock fronts propagating into surrounding ISM and
accelerating cosmic rays, or they result from an interaction of ejecta with the envelope of the progenitor SNR, presumably a Type IIP supernova \citep{bocchetal2005}.
Since the North Spur and Eastern Limb have different spectral and spatial properties, we discuss them separately.  

A multi-wavelength analysis is required to get a full understanding of the North Spur.  This feature was observed in the radio and the soft X-ray, most 
recently by \citet{beit2011} and \citet{msh2010}, respectively.
\citeauthor{beit2011} reported a radio detection of the North Spur, with a 1.43 GHz flux density of $20.2\pm1.8$~mJy, and a FWHM size of $18''\times8''$.  
This is notable because it is the only feature, other than the PWN itself that is detected in the radio band.  The Eastern Limb has no radio emission detected
to date.
 
\citet{msh2010} obtained $\sim580$~ks of \chandra\ data, and found that the North Spur has a spectrum comprised of non-thermal 
and thermal components. The thermal component, 
represented by the {\bfseries pshock} model, is best-fit to temperature of $kT\sim0.2$~keV and contributes only $\sim6-7\%$ to 
the overall $0.5-8$~keV flux.  The non-thermal component, however, is equally well-fit by either the {\bfseries srcut} or {\bfseries powerlaw} model.  

The model {\bfseries srcut} describes the synchrotron emission from a homogeneous source consisting of a power-law energy distribution of electrons 
with an exponential cutoff, radiating in a uniform magnetic field.  The emitted spectrum is a power-law that steepens slowly above the photon energy 
corresponding to the electron cutoff energy.  This slow curvature can mimic a steeper power-law in a limited energy band. However, in a broader 
energy band the {\bfseries srcut} model can fall well below the extrapolation of a power-law with the same slope at lower energies.
Since X-ray emission is visible from the North Spur at energies as high as $20$~keV, the correct spectral model 
must provide a photon flux from $15-20$~keV that is statistically higher than the background.  

Due to signal-to-noise limitations we can not use \nustar\ data to spectrally fit the Spur, however 
we can use the deconvolved images to distinguish between the {\bfseries srcut} and {\bfseries powerlaw} models.
We simulated spectra with these models using the {\it fakeit} command in XSPEC using the model parameters reported by \citet{msh2010}.
A photon spectral index of $\Gamma=2.21$ was applied to the {\bfseries powerlaw} model, while the {\bfseries srcut} model used a radio index of $\alpha=0.8$ and 
a rolloff frequency of $\nu_{\text{rolloff}}=18\times10^{17}$~Hz.
The \nustar\ response files were based on a point source extraction with $r=30''$.  We were thus able to obtain 
fluxes from each respective model from $10-15$~keV, $15-20$~keV, and from $20-25$~keV.

Both models have count rates higher than that of the background within the $10-15$~keV energy band.  In the $15-20$~keV band, the {\bfseries powerlaw} model
has a count rate 7 times higher than the background, while the {\bfseries srcut} model is only 3 times higher.  Finally, within the $15-20$~keV band,
the count rates are 4 and 1.5 higher than the background for the {\bfseries powerlaw} and {\bfseries srcut} models, respectively. This implies that
the {\bfseries srcut} model should be marginally visible up to $20$~keV, and above $20$~keV should have a count rate equal to that of
the background. This matches well with what is seen in the \nustar\ images.  The {\bfseries powerlaw} model, however, should be detectable at energies as high 
as $25$~keV.

If the North Spur were an extension of the PWN, it would have a spectral photon index similar to that of the PWN itself.  While this is true, 
the analysis above indicates that the North Spur is not described by a {\bfseries powerlaw} model extending to higher energies.  If it were, the North Spur
would easily be detectable at energies as high as $25$~keV.  However, the \nustar\ images with a combined exposure of 281 ksec do not detect any statistical 
emission above $20$~keV, indicating that {\bfseries srcut} is the more plausible spectral model for the North Spur.

It is possible that the North Spur results from the interaction of the inner SN ejecta with the H-envelope of the progenitor \citep{bocchetal2005, msh2010}.
This is supported by the thermal component in the spectral fit of \citet{msh2010}, a {\bfseries pshock} model with solar abundances, low temperature of $kT\sim0.21\pm0.4$~keV,
and low ionization timescales.  This is also supported by the morphology of the North Spur itself.  With the inclusion of projection effects, the North Spur 
is located between $\sim75''$ and the edge of the SN shell at $\sim120''$.

\subsection{The Eastern Limb: the shell of \g21}

X-ray emission from the shell of \g21\ is clearly visible in the \nustar\ image. Our results confirm the existence of this shell, revealed 
by \chandra\ \citep{msh2010} in both their image and in an extracted shell spectrum (up to $\sim6$~keV), but also hinted at in earlier \XMM\ data \citep{bocchetal2005}. 
The \XMM\ observations reveal evidence of the shell in a $\sim2-8$ keV energy band image, after careful subtraction of a modeled 
dust scattering component below $\sim5$ keV.  The \nustar\ detection extends up to much a higher energy of $\sim20$ keV. The morphology of the \nustar\ emission is striking 
in its similarity to the \chandra\ and especially the \XMM\ image. Emission is detected from PA $\sim180^{\circ}$ to $\sim300^{\circ}$ in the $6-10$~keV image 
(PA = 0 at N, positive clockwise), with the extent 
shrinking as the energy increases until it is visible mainly in the east and north at the highest energies. This is consistent with the intensity 
distribution with position angle seen in the lower energy image.  

A question unresolved by previous X-ray observations is whether the shell emission is thermal or non-thermal.  The extended energy response of \nustar\ can be 
exploited to answer this question.  \citet{msh2010} found that the spectrum of the Eastern limb could be characterized equally by four distinct models. 
A thermal fit to the {\bfseries pshock} model provided a temperature of $kT\sim7.5$~keV while a non-thermal {\bfseries powerlaw} model obtained a spectral photon index of 
$\Gamma=2.13$.  Two {\bfseries srcut} models were also well-fit to the Eastern Limb spectrum.  Ideally {\bfseries srcut} requires both a radio flux density and radio 
spectral index for the shell, but the Eastern Limb has not been detected
in the radio.  Therefore, two values of the radio index $\alpha$ that are reasonable for a SN shell were chosen ($\alpha=0.3/0.5$) while all
other parameters were left free to vary. Care was taken to ensure the best-fit radio flux was below the upper limit reported by \citet{beit2011}. 

As with the North Spur, we extrapolated the four \chandra\ spectra for the Eastern Limb into the \nustar\ band. We created an effective area file for 
an extended source shaped like the Eastern Limb, then used the {\it fakeit} command in XSPEC to simulate spectral data. 
The thermal {\bfseries pshock} model predicted a shell flux which would not produce the
X-ray emission seen by \nustar\ at $\sim15-20$~keV.  In contrast, the three non-thermal models produced X-ray fluxes consistent with imaging of the 
Eastern Limb by \nustar, although the {\bfseries srcut} and {\bfseries powerlaw} models could not be distinguished from each other.  Nonetheless,
the \nustar\ observations firmly establish the non-thermal nature of the shell X-ray emission.

The detection of a non-thermal shell up to quite high X-ray energy in \g21\ is interesting, and is in contrast to observations of other Crab-like supernova remnants.
Recently shells have been detected in 3C58 \citep{gott2007} and G54.1+0.3 \citep{bbg2010}, and a clear, detached shell of emission in Kes 75 \citep{hcg2003}.
However the 3C58 shell is clearly thermal, with no sign of a non-thermal component. The shell of ${\rm G}54.1+0.3$ can be fit with both thermal and non-thermal 
models, however a thermally emitting shell seems much more likely. Assuming thermal emission, \citet{bbg2010} were able to use PWN-SNR evolutionary 
models to obtain an age consistent with the characteristic age derived from pulsar observations, obtain the proper supernova-PWN radius ratio, and predict 
that the reverse shock has not encountered the PWN yet, consistent with other observations. The ages of 3C58 and G54.1+0.3 are $\sim3000-5000$ years and 
$\sim1800-3300$ years, respectively.  \g21\ is thus unique in that it is much younger ($\sim290-1000$ years), and potentially has higher forward 
shock speed, both of which could lead to the detectable non-thermal shell.  The Crab Nebula itself, of comparable age to \g21\ but much closer, still shows no 
non-thermal shell, presumably due to a very low interstellar medium density. The detection of this \g21\ non-thermal shell at quite high 
X-ray energy is thus likely due to its younger age compared to these other Crab-like SNRs.

\section {Summary}

We have presented an analysis of a 281~ks  \nustar\ observation of \g21\ to probe the spatial and spectral characteristics 
revealed by high-energy X-ray emission.  \nustar's broad energy band 
makes it uniquely suited to analyze not only the PWN but observe the characteristics of the SNR shell as well.
The spectrum taken from the entire remnant is described by a best-fit broken power-law with a break energy of $\sim9$~keV.  This is the first instance where
a single instrument was able to capture this break.  
The addition of the \nustar\ spectrum to SED models from \citet{tt2011} produces a poor fit to the X-ray data.  This suggests that further modeling is required: 
more complex electron injection spectra, additional loss processes such as diffusion, or radial dependence of the PWN parameters. 

Spectra extracted from various radial annuli were also fit with both an absorbed power-law and an absorbed broken power-law model.  
The two regions $r\leq30''$ and $r=30-60''$ are statistically better fit with a broken power law with $E_{\rm break}\sim9$~keV, while the regions with 
radii $r>60''$ are best-fit with a single power law.  
We observe spectral softening of the spectral index below the spectral break, while the spectral index above the break is constant within uncertainties.

Image analysis allows us to measure the energy-dependent cooling length scale and fit the relationship with a power-law model of $L(E)\propto E^{m}$. 
This yields an index of $m=-0.21\pm0.01$.
Incorporating this with the spectral indices in both the radio and X-ray bands, we are able to systematically approach the equations of \citet{rey2009} and 
inspect the parameter space for physically consistent solutions.  We found that, for a conical jet or spherical outflow, the most reasonable solutions do not conserve magnetic flux but do conserve mass, 
indicating turbulent magnetic field amplification.  The bulk velocity decelerates steeper than that predicted by KC84.

We detect the Eastern Limb and North Spur at energies above 10~keV for the first time. 
A deconvolution method provids clear evidence of emission from the North Spur up to $20$~keV.  Extrapolation of the 
spectral fits obtained by \chandra\ show that the {\bfseries srcut} model is favored over the {\bfseries powerlaw} model.  This further solidifies the assumption that the 
North Spur is an interaction of the SN ejecta with the remnant.
We also detect the Eastern Limb up to $20$~keV.  We have confirmed the existence of non-thermal emission from the Limb, and conclude this faint feature is the
SN shell of \g21.  We are unable to distinguish between the non-thermal models fit to the Eastern Limb by \chandra.

\acknowledgements
This work was supported under NASA Contract No. NNG08FD60C, and made use of data from the \nustar\ mission, a project led 
by the California Institute of Technology, managed by the Jet Propulsion Laboratory, and funded by the National Aeronautics and Space Administration.
We thank the \nustar\ Operations, Software and Calibration teams for support with the execution and analysis of these observations. This research has 
made use of the \nustar\ Data Analysis Software (NuSTARDAS) jointly developed by the ASI Science Data Center (ASDC, Italy) and the California Institute
of Technology (USA).


\begin{thebibliography}{}
\expandafter\ifx\csname natexlab\endcsname\relax\def\natexlab#1{#1}\fi

\bibitem[{{Abdo} {et~al.}(2010){Abdo}, {Ackermann}, {Ajello}, {Atwood},
  {Axelsson}, {Baldini}, {Ballet}, {Barbiellini}, {Baring}, {Bastieri}, \&
  et~al.}]{abdo2010}
{Abdo}, A.~A., {Ackermann}, M., {Ajello}, M., {et~al.} 2010, \apjs, 187, 460

\bibitem[{{Ackermann} {et~al.}(2011){Ackermann}, {Ajello}, {Baldini}, {Ballet},
  {Barbiellini}, {Bastieri}, {Bechtol}, {Bellazzini}, {Berenji}, {Bloom},
  {Borgland}, {Bouvier}, {Bregeon}, {Brez}, \& {Brigida}}]{Ack2011}
{Ackermann}, M., {Ajello}, M., {Baldini}, L., {et~al.} 2011, \apj, 726, 35

\bibitem[{{Altenhoff} {et~al.}(1970){Altenhoff}, {Downes}, {Goad}, {Maxwell},
  \& {Rinehart}}]{alt1970}
{Altenhoff}, W.~J., {Downes}, D., {Goad}, L., {Maxwell}, A., \& {Rinehart}, R.
  1970, \aaps, 1, 319

\bibitem[{{Arnaud}(1996)}]{arnaud96}
{Arnaud}, K.~A. 1996, in ASP conf., Vol. 101, Astronomical Data Analysis
  Software and Systems V, ed. G.~H. {Jacoby} \& J.~{Barnes}, 17

\bibitem[{{Bandiera} \& {Bocchino}(2004)}]{bandbocc2004}
{Bandiera}, R., \& {Bocchino}, F. 2004, Advances in Space Research, 33, 398

\bibitem[{{Bandiera} {et~al.}(2001){Bandiera}, {Neri}, \&
  {Cesaroni}}]{Bandiera2001}
{Bandiera}, R., {Neri}, R., \& {Cesaroni}, R. 2001, in American Institute of
  Physics Conference Series, Vol. 565, Young Supernova Remnants, ed. S.~S.
  {Holt} \& U.~{Hwang}, 329--332

\bibitem[{{Becker} \& {Kundu}(1976)}]{bk1976}
{Becker}, R.~H., \& {Kundu}, M.~R. 1976, \apj, 204, 427

\bibitem[{{Becker} \& {Szymkowiak}(1981)}]{bs1981}
{Becker}, R.~H., \& {Szymkowiak}, A.~E. 1981, \apjl, 248, L23

\bibitem[{{Bietenholz} {et~al.}(2011){Bietenholz}, {Matheson}, {Safi-Harb},
  {Brogan}, \& {Bartel}}]{beit2011}
{Bietenholz}, M.~F., {Matheson}, H., {Safi-Harb}, S., {Brogan}, C., \&
  {Bartel}, N. 2011, \mnras, 412, 1221

\bibitem[{{Bocchino}(2005)}]{bocch2005}
{Bocchino}, F. 2005, Advances in Space Research, 35, 1003

\bibitem[{{Bocchino} {et~al.}(2010){Bocchino}, {Bandiera}, \&
  {Gelfand}}]{bbg2010}
{Bocchino}, F., {Bandiera}, R., \& {Gelfand}, J. 2010, \aap, 520, A71

\bibitem[{{Bocchino} \& {Bykov}(2001)}]{bb2001}
{Bocchino}, F., \& {Bykov}, A.~M. 2001, \aap, 376, 248

\bibitem[{{Bocchino} {et~al.}(2005){Bocchino}, {van der Swaluw}, {Chevalier},
  \& {Bandiera}}]{bocchetal2005}
{Bocchino}, F., {van der Swaluw}, E., {Chevalier}, R., \& {Bandiera}, R. 2005,
  \aap, 442, 539

\bibitem[{{Camilo} {et~al.}(2006){Camilo}, {Ransom}, {Gaensler}, {Slane},
  {Lorimer}, {Reynolds}, {Manchester}, \& {Murray}}]{cam2006}
{Camilo}, F., {Ransom}, S.~M., {Gaensler}, B.~M., {et~al.} 2006, \apj, 637, 456

\bibitem[{{Chevalier}(2005)}]{chev2005}
{Chevalier}, R.~A. 2005, \apj, 619, 839

\bibitem[{{de Jager} {et~al.}(2008){de Jager}, {Ferreira}, \&
  {Djannati-Ata{\"i}}}]{dj2008}
{de Jager}, O.~C., {Ferreira}, S.~E.~S., \& {Djannati-Ata{\"i}}, A. 2008, in
  American Institute of Physics Conference Series, Vol. 1085, American
  Institute of Physics Conference Series, ed. F.~A. {Aharonian}, W.~{Hofmann},
  \& F.~{Rieger}, 199--202

\bibitem[{{Fruscione} {et~al.}(2006){Fruscione}, {McDowell}, {Allen},
  {Brickhouse}, {Burke}, {Davis}, {Durham}, {Elvis}, {Galle}, {Harris},
  {Huenemoerder}, {Houck}, {Ishibashi}, {Karovska}, {Nicastro}, {Noble},
  {Nowak}, {Primini}, {Siemiginowska}, {Smith}, \& {Wise}}]{frusc2006}
{Fruscione}, A., {McDowell}, J.~C., {Allen}, G.~E., {et~al.} 2006, in Society
  of Photo-Optical Instrumentation Engineers (SPIE) Conference Series, Vol.
  6270, Society of Photo-Optical Instrumentation Engineers (SPIE) Conference
  Series

\bibitem[{{Furst} {et~al.}(1998){Furst}, {Reich}, {Uyaniker}, \&
  {Wielebinski}}]{furst1998}
{Furst}, E., {Reich}, W., {Uyaniker}, B., \& {Wielebinski}, R. 1998, in IAU
  Symposium, Vol. 179, New Horizons from Multi-Wavelength Sky Surveys, ed.
  B.~J. {McLean}, D.~A. {Golombek}, J.~J.~E. {Hayes}, \& H.~E. {Payne}, 97

\bibitem[{{Gaensler} \& {Slane}(2006)}]{gs2006}
{Gaensler}, B.~M., \& {Slane}, P.~O. 2006, \araa, 44, 17

\bibitem[{{Goss} \& {Day}(1970)}]{gd1970}
{Goss}, W.~M., \& {Day}, G.~A. 1970, Australian Journal of Physics
  Astrophysical Supplement, 13, 3

\bibitem[{{Gotthelf} {et~al.}(2007){Gotthelf}, {Helfand}, \&
  {Newburgh}}]{gott2007}
{Gotthelf}, E.~V., {Helfand}, D.~J., \& {Newburgh}, L. 2007, \apj, 654, 267

\bibitem[{{Gratton}(1972)}]{gratton1972}
{Gratton}, L. 1972, \apss, 16, 81

\bibitem[{{Gupta} {et~al.}(2005){Gupta}, {Mitra}, {Green}, \&
  {Acharyya}}]{gup2005}
{Gupta}, Y., {Mitra}, D., {Green}, D.~A., \& {Acharyya}, A. 2005, Current
  Science, 89, 853

\bibitem[{{Harrison} {et~al.}(2013){Harrison}, {Craig}, {Christensen},
  {Hailey}, {Zhang}, {Boggs}, \& {Stern}}]{har2013}
{Harrison}, F.~A., {Craig}, W.~W., {Christensen}, F.~E., {et~al.} 2013, \apj,
  770, 103

\bibitem[{{Helfand} {et~al.}(2003){Helfand}, {Collins}, \&
  {Gotthelf}}]{hcg2003}
{Helfand}, D.~J., {Collins}, B.~F., \& {Gotthelf}, E.~V. 2003, \apj, 582, 783

\bibitem[{{Kennel} \& {Coroniti}(1984)}]{kc84a}
{Kennel}, C.~F., \& {Coroniti}, F.~V. 1984, \apj, 283, 694

\bibitem[{{Komissarov} \& {Lyubarsky}(2004)}]{kl2004}
{Komissarov}, S.~S., \& {Lyubarsky}, Y.~E. 2004, \mnras, 349, 779

\bibitem[{{Ku} {et~al.}(1976){Ku}, {Kestenbaum}, {Novick}, \& {Wolff}}]{ku1976}
{Ku}, W., {Kestenbaum}, H.~L., {Novick}, R., \& {Wolff}, R.~S. 1976, \apjl,
  204, L77

\bibitem[{{Lucy}(1974)}]{lucy1974}
{Lucy}, L.~B. 1974, \apj, 79, 745

\bibitem[{{Madsen}(2014)}]{madsen2014}
{Madsen}, K.~C. 2014, \apj, in preparation

\bibitem[{{Massaro}(1985)}]{massaro1985}
{Massaro}, E. 1985, \apss, 108, 369

\bibitem[{{Matheson} \& {Safi-Harb}(2005)}]{msh2005}
{Matheson}, H., \& {Safi-Harb}, S. 2005, Advances in Space Research, 35, 1099

\bibitem[{{Matheson} \& {Safi-Harb}(2010)}]{msh2010}
---. 2010, \apj, 724, 572

\bibitem[{{Predehl} \& {Schmitt}(1995)}]{ps1995}
{Predehl}, P., \& {Schmitt}, J.~H.~M.~M. 1995, \aap, 293, 889

\bibitem[{{Rees} \& {Gunn}(1974)}]{rg1974}
{Rees}, M.~J., \& {Gunn}, J.~E. 1974, \mnras, 167, 1

\bibitem[{{Reynolds}(2003)}]{rey2003}
{Reynolds}, S.~P. 2003, ArXiv Astrophysics e-prints, arXiv:astro-ph/0308483

\bibitem[{{Reynolds}(2009)}]{rey2009}
---. 2009, \apj, 703, 662

\bibitem[{{Richardson}(1972)}]{rich1972}
{Richardson}, W.~H. 1972, Journal of the Optical Society of America
  (1917-1983), 62, 55

\bibitem[{{Safi-Harb} {et~al.}(2001){Safi-Harb}, {Harrus}, {Petre}, {Pavlov},
  {Koptsevich}, \& {Sanwal}}]{sh2001}
{Safi-Harb}, S., {Harrus}, I.~M., {Petre}, R., {et~al.} 2001, \apj, 561, 308

\bibitem[{{Salter} {et~al.}(1989){Salter}, {Emerson}, {Steppe}, \&
  {Thum}}]{salter1989}
{Salter}, C.~J., {Emerson}, D.~T., {Steppe}, H., \& {Thum}, C. 1989, \aap, 225,
  167

\bibitem[{{Slane} {et~al.}(2000){Slane}, {Chen}, {Schulz}, {Seward}, {Hughes},
  \& {Gaensler}}]{slane2000}
{Slane}, P., {Chen}, Y., {Schulz}, N.~S., {et~al.} 2000, \apjl, 533, L29

\bibitem[{{Smith}(2008)}]{smith2008}
{Smith}, R.~K. 2008, \apj, 681, 343

\bibitem[{{Smith} {et~al.}(2002){Smith}, {Edgar}, \& {Shafer}}]{smith2002}
{Smith}, R.~K., {Edgar}, R.~J., \& {Shafer}, R.~A. 2002, \apj, 581, 562

\bibitem[{{Tanaka} \& {Takahara}(2011)}]{tt2011}
{Tanaka}, S.~J., \& {Takahara}, F. 2011, \apj, 741, 40

\bibitem[{{Tang} \& {Chevalier}(2012)}]{tc2012}
{Tang}, X., \& {Chevalier}, R.~A. 2012, \apj, 752, 83

\bibitem[{{Tian} \& {Leahy}(2008)}]{tl2008}
{Tian}, W.~W., \& {Leahy}, D.~A. 2008, \mnras, 391, L54

\bibitem[{{Tsujimoto} {et~al.}(2011){Tsujimoto}, {Guainazzi}, {Plucinsky},
  {Beardmore}, {Ishida}, {Natalucci}, {Posson-Brown}, {Read}, {Saxton}, \&
  {Shaposhnikov}}]{tsuj2011}
{Tsujimoto}, M., {Guainazzi}, M., {Plucinsky}, P.~P., {et~al.} 2011, \aap, 525,
  A25

\bibitem[{{Verner} {et~al.}(1996){Verner}, {Ferland}, {Korista}, \&
  {Yakovlev}}]{vern96}
{Verner}, D.~A., {Ferland}, G.~J., {Korista}, K.~T., \& {Yakovlev}, D.~G. 1996,
  \apj, 465, 487

\bibitem[{{Vorster} {et~al.}(2013){Vorster}, {Tibolla}, {Ferreira}, \&
  {Kaufmann}}]{vorster2013}
{Vorster}, M.~J., {Tibolla}, O., {Ferreira}, S.~E.~S., \& {Kaufmann}, S. 2013,
  \apj, 773, 139

\bibitem[{{Warwick} {et~al.}(2001){Warwick}, {Bernard}, {Bocchino},
  {Decourchelle}, {Ferrando}, {Griffiths}, {Haberl}, {La Palombara}, {Lumb},
  {Mereghetti}, {Read}, {Schaudel}, {Schurch}, {Tiengo}, \&
  {Willingale}}]{war2001}
{Warwick}, R.~S., {Bernard}, J.-P., {Bocchino}, F., {et~al.} 2001, \aap, 365,
  L248

\bibitem[{{Wilms} {et~al.}(2000){Wilms}, {Allen}, \& {McCray}}]{wilms2000}
{Wilms}, J., {Allen}, A., \& {McCray}, R. 2000, \apj, 542, 914

\bibitem[{{Wilson}(1972)}]{wilson1972}
{Wilson}, A.~S. 1972, \mnras, 160, 355

\bibitem[{{Wilson} \& {Altenhoff}(1970)}]{wa1970}
{Wilson}, T.~L., \& {Altenhoff}, W. 1970, \aplett, 5, 47

\end{thebibliography}
\end{document}